\documentclass[a4paper,reqno,10pt]{amsart}

\usepackage[
    a4paper,
    top=4cm,
    bottom=4cm,
    left=3cm,
    right=3cm
]{geometry}

\usepackage{indentfirst}
\usepackage[normalem]{ulem}

\usepackage[dvipsnames]{xcolor}
\usepackage{microtype}

\usepackage{newpxtext}

\usepackage{newpxmath}
\usepackage[mathcal]{eucal}
\usepackage{mathrsfs}
\usepackage{mathtools}

\usepackage{graphicx}
\usepackage[shortlabels]{enumitem}
\usepackage{comment}

\usepackage[
    backend=biber,
    style=alphabetic,
    sorting=nyt,
    maxalphanames=4,
    maxbibnames=99,
    maxcitenames=2
]{biblatex}

\usepackage{thmtools}
\usepackage{hyperref}

\hypersetup{
    colorlinks=true,
    allcolors=Maroon
}

\usepackage[nameinlink,noabbrev]{cleveref}

\makeatletter

\def\section{\@startsection{section}{1}%
    \z@
    {2.5ex \@plus 1ex \@minus .2ex}%
    {2ex}%
    {\normalfont\Large\bfseries\raggedright}}

\def\subsection{\@startsection{subsection}{2}%
    \z@
    {2.3ex \@plus 1ex \@minus .2ex}%
    {2ex}%
    {\normalfont\large\bfseries\raggedright}}

\def\@settitle{%
    \begin{center}%
        {\normalfont\bfseries\LARGE\@title}%
    \end{center}%
}

\makeatother

\theoremstyle{definition}
\newtheorem{bump}{Bump}[section]

\theoremstyle{plain}
\newtheorem{theorem}[bump]{Theorem}
\newtheorem{proposition}[bump]{Proposition}
\newtheorem{definition}[bump]{Definition}
\newtheorem{lemma}[bump]{Lemma}
\newtheorem{corollary}[bump]{Corollary}

\newtheorem{remark}[bump]{Remark}

\RenewCommandCopy{\theHbump}{\thebump}
\RenewCommandCopy{\theHtheorem}{\thetheorem}
\RenewCommandCopy{\theHproposition}{\theproposition}
\RenewCommandCopy{\theHdefinition}{\thedefinition}
\RenewCommandCopy{\theHlemma}{\thelemma}
\RenewCommandCopy{\theHcorollary}{\thecorollary}
\RenewCommandCopy{\theHhypothesis}{\thehypothesis}
\RenewCommandCopy{\theHconjecture}{\theconjecture}
\RenewCommandCopy{\theHremark}{\theremark}
\RenewCommandCopy{\theHexample}{\theexample}
\RenewCommandCopy{\theHnotation}{\thenotation}

\usepackage{etoolbox}

\makeatletter

\renewcommand{\email}[2][]{%
    \g@addto@macro\addresses{%
        \email{#1}{\href{mailto:#2}{#2}}%
    }%
}

\pretocmd{\@setaddresses}
    {\vspace{1.5\baselineskip}}
    {}{}

\makeatother

\newcommand{\redmargin}[1]{%
    \marginpar{%
        \raggedright
        \footnotesize
        \color{red}#1%
    }%
}

\newcommand{\printtitledate}[1]{%
    \vspace*{-0.6\baselineskip}%
    \begin{center}
        \normalfont\small #1
    \end{center}%
    \vspace*{-0.3\baselineskip}%
}

\begin{document}

\title[$C^0$-inextendibility of a class of Vaidya spacetimes]{ $C^0$- inextendibility of a class of  Vaidya spacetimes}

\author[Mosani]{Karim Mosani}
\email{karim.mosani@univie.ac.at}
\author[Sämann]{Clemens Sämann}
\email{clemens.saemann@univie.ac.at}
\author[Solanki]{Kharanshu N. Solanki}
\email{kharanshu.nilesh.solanki@univie.ac.at}

\address{Fakultät für Mathematik, Universität Wien,  Oskar-Morgenstern-Platz 1, 1090 Wien, Austria}


\maketitle

\printtitledate{\today}

\begin{abstract}  

We prove the future $C^0$-inextendibility of a broad subclass of ingoing Vaidya spacetimes with positive mass function satisfying the weak energy condition, building on methods introduced and developed by Sbierski for the Schwarzschild spacetime.

\end{abstract}
\thispagestyle{empty}

\tableofcontents

\addtocontents{toc}{\protect\setcounter{tocdepth}{2}}

\section{Introduction }

A fundamental problem in the study of spacetime singularities is to determine whether a given spacetime admits an extension as a Lorentzian manifold equipped with a low-regularity metric tensor. The strong cosmic censorship conjecture asserts, in one of its standard formulations, that for generic compact or asymptotically flat initial data for Einstein's equations, in vacuum or coupled to suitable matter fields, the corresponding maximal globally hyperbolic development (MGHD) is inextendible as a Lorentzian manifold \emph{within a prescribed regularity class} \cite{VanDeMoortel2025SCC}. That the MGHD is inextendible \emph{within the class of globally hyperbolic developments that solve the Einstein equations with the same initial data}, is precisely the conclusion of the well-established global existence result for MGHDs \cite{Choquet-Bruhat:1969ywq,Sbierski:2013kca}. 

At the $C^2$ level, unbounded curvature 
provides a direct mechanism for proving inextendibility. The Schwarzschild spacetime is the basic example: its Kretschmann scalar diverges as the radial coordinate approaches zero, ruling out any $C^2$ extension across the central singularity. At the level of continuous metrics, however, curvature is no longer available as a directly meaningful quantity, and curvature blow-up by itself need not prevent extendibility. Indeed, there are spacetimes that are $C^2$-inextendible while still admitting $C^0$-extensions (see, for e.g., Section 1.1 in \cite{Sbierski:2015nta}). This motivates the search for geometric mechanisms of proving inextendibility that remain meaningful at the level of continuous metrics.

More precisely, a \emph{$C^0$-extension} of a Lorentzian manifold $(M,g)$ is a Lorentzian manifold $(\tilde M,\tilde g)$, with $\tilde g$ continuous, together with an isometric embedding $\iota:M\hookrightarrow \tilde M$ such that $\iota(M)\subsetneq \tilde M$ and $\operatorname{dim} M=\operatorname{dim}\tilde M$. If no such extension exists, $(M,g)$ is said to be \emph{$C^0$-inextendible}. 

Suppose, in addition, that $(M,g)$ is time-oriented. Given a $C^0$-extension $\iota: M \hookrightarrow\tilde M$, a point $p\in\partial\iota(M)$ is called a \emph{future boundary point} of $M$ if there exists a timelike curve $\tilde{\gamma}:[-1,0] \to \tilde{M}$ such that $\operatorname{Im}(\tilde{\gamma}|_{[-1,0)}) \subseteq \iota (M)$, $\tilde{\gamma}(0) = p$, and  $\iota^{-1}\circ (\tilde{\gamma}|_{[-1,0)})$ is future-directed in  $M$.  We call the set of all such points the \emph{future boundary} of $M$ and denote it by $\partial^+\iota(M).$ Finally, a \emph{future $C^0$-extension} of a time-oriented Lorentzian manifold $(M,g)$ is a $C^0$-extension $\iota:M\hookrightarrow\tilde M$  with  $\partial^+\iota(M)\neq \emptyset$. Accordingly, a time-oriented Lorentzian manifold $(M,g)$ is said to be \emph{future $C^0$-inextendible} if it admits no future $C^0$-extension.

The gap between curvature blow-up and $C^0$-inextendibility becomes even more pronounced when one considers stronger notions of spacetime singularities.
Tipler introduced a notion of strong singularities in terms of the vanishing of volume elements generated by suitable Jacobi fields along causal geodesics terminating at such singularities \cite{Tipler:1977zza,Tipler:1977zzb}. A sufficient condition for this behaviour is a lower bound on the rate of blow-up of Ricci curvature along these curves, and this has the physical interpretation that an extended object approaching the singularity is crushed to zero volume. Motivated by this picture, Tipler conjectured that a spacetime would be $C^0$-inextendible across such \emph{strong curvature singularities}. This intuition, however, is not valid in complete generality. Ori constructed an example of a spacetime possessing a strong curvature singularity in the sense of Tipler while nevertheless admitting a $C^0$-extension \cite{PhysRevD.61.064016} (see also the preceding work by \cite{Nolan:1999tw} in which he emphasised that the presence of a strong singularity in the sense of Tipler does not, in principle, rule out the existence of a $C^0$-extension of the spacetime). Such an example serves as further evidence that neither unbounded curvature nor its rate of blow-up can, by itself, serve as a general criterion for $C^0$-inextendibility.

A major development in the study of $C^0$-inextendibility is Sbierski's proof that the maximal analytic Schwarzschild spacetime, i.e. the Kruskal extension, is $C^0$-inextendible \cite{Sbierski:2015nta,Sbierski_2018}. 
Subsequent work extended the study of $C^0$-extendibility and inextendibility to a variety of cosmological as well as black hole settings.

In the cosmological setup, Galloway and Ling studied FLRW spacetimes in \cite{Galloway:2016bej}, obtaining, under suitable assumptions on the scale factor, both extendibility and inextendibility results for spatially flat and hyperbolic FLRW models. In particular, they constructed continuous extensions of Milne-like spacetimes through the big bang, while establishing inextendibility of suitable spatially flat and non-Milne-like hyperbolic models within the class of strongly spherically symmetric extensions. Without imposing any symmetry conditions on the extension, Sbierski proved past $C^0$-inextendibility for spherical FLRW spacetimes without particle horizons, as well as hyperbolic FLRW spacetimes without particle horizon with an additional assumption on the scale factor  \cite{Sbierski:2023tbc}. Additionally, Graf and van den Beld-Serrano generalised the earlier FLRW results of \cite{Galloway:2016bej} to a broader class of axisymmetric extensions \cite{Graf:2024wpy}. Ling later establish past $C^0$-inextendibility for a class of past-eternal spatially flat FLRW spacetimes \cite{Ling:2024nuc}, and Miethke proved $C^0$-inextendibility of a subclass of Kasner spacetimes \cite{Miethke:2024bfm}. 

In the black hole setup, $C^0$-inextendibility has recently been established for a broad class of warped-product black hole spacetimes with static exterior regions and without the assumption of spherical symmetry \cite{Mosani_2026}. It is natural to ask whether analogous $C^0$-inextendibility results hold in genuinely dynamical black hole settings. The Vaidya spacetime \cite{Vaidya1999,Vaidya1953} provides a natural model in which to investigate this question. It is a spherically symmetric spacetime that, in certain cases, describes the gravitational collapse of null radiation and may be viewed as a dynamical generalisation of the Schwarzschild spacetime. In specific coordinates called Eddington--Finkelstein coordinates, its metric has the same form as the Schwarzschild metric, with the constant Schwarzschild mass replaced by a mass function $m(v)$ depending on the advanced null coordinate $v$. 

\subsection{Main Result}

We first define the class of Vaidya spacetimes that we deal with here, and then mention and interpret the main result. For further background on Vaidya spacetimes, we refer to Sections~3.5 and~6.5 of \cite{10.1093/oso/9780198539667.001.0001} and Section~9.5 of \cite{Griffiths:2009dfa}. 
\begin{definition}[Ingoing Vaidya spacetimes]\label{Vaidya-spacetime}
 Let $ I\subseteq\mathbb{R}$ be a nonempty open interval and consider the smooth manifold $M\coloneqq  I\times (0,\infty) \times \mathbb{S}^2$. We denote by $(v,r)$ the global coordinates on $I \times (0,\infty)$ and denote the spherical coordinates by $\omega$. We equip $M$ with the Lorentzian metric,
\begin{equation}\label{vaidya-original-metric}
    g = -\mathcal{F}(v,r) \ dv \otimes dv + (dr\otimes dv + dv\otimes dr) +r^2 g_{\mathbb{S}^2}, \quad \mathcal F(v,r)\coloneqq 1-\frac{2m(v)}{r},
\end{equation}
where $m:{I} \to (0,\infty)$ is a $C^2$ function called the \textit{mass function}, and $g_{\mathbb{S}^2}$ is the standard round metric on the sphere. A time orientation on $(M,g)$ is fixed by declaring $-\partial_r$ to be future-directed null. The pair $(M,g)$ is called an ingoing Vaidya spacetime. 
\end{definition}

The ingoing Vaidya spacetime is a spherically symmetric solution of the Einstein field equations sourced by infalling radiation or \emph{null dust}, which is given by the energy-momentum tensor
\begin{equation*}
    T = \frac{m'(v)}{r^2} dv \otimes dv,
\end{equation*}
where $m'(v) \coloneqq dm/dv$ and $m'(v)/r^2$ can be interpreted as the energy density of radiation. It follows that the \emph{weak energy condition} $T(X,X)\geq 0$ for timelike $X$ is equivalent to $m'(v)\geq 0$.

There is a corresponding \emph{outgoing Vaidya spacetime} obtained by reversing the time orientation. Let $J:=-I$ and $\bar m(u):=m(-u)$, and set $M_{\mathrm{o}}=J\times(0,\infty)\times S^2$. The outgoing Vaidya metric is 

\begin{equation}
\label{outgoing-metric}
    g_{\mathrm{o}} =-\left(1-\frac{2\bar m(u)}r\right)du\otimes du-\bigl(dr\otimes du+du\otimes dr\bigr)+r^2g_{S^2}.
\end{equation}

Its time orientation is fixed by declaring $\partial_r$ to be future-directed null. The map $\mathcal T:M_{\mathrm{o}}\to M$, defined by $ \mathcal T(u,r,\omega)\coloneqq (-u,r,\omega),$ is a time-orientation-reversing isometry. Hence outgoing Vaidya spacetimes are precisely the time reversals of ingoing Vaidya spacetimes.\\

The Kretschmann scalar for the ingoing Vaidya spacetimes is given by $R^{\mu\nu\alpha\beta}R_{\mu\nu\alpha\beta}=48m(v)^2/r^6$. For positive non-decreasing mass functions, we have that $R^{\mu\nu\alpha\beta}R_{\mu\nu\alpha\beta}\to \infty$ along timelike curves for which $r\to0$ at their future endpoints. Consequently, the positive mass ingoing Vaidya spacetimes satisfying the weak energy condition are $C^2$-inextendible across the \emph{central singularity} $\{r=0\}$ (see Lemma 18.18 in \cite{Ringstrom2009}). The main result of this paper substantially strengthens this statement to the following.


\begin{theorem}[Ingoing Vaidya spacetime is future $C^0$-inextendible]\label{thm:main thm3}
     Consider an ingoing Vaidya spacetime $(M,g)$, as given in \Cref{Vaidya-spacetime} such that it satisfies the weak energy condition $m'(v)\geq 0$ and the non-integrability condition 
     \begin{equation}
         \label{non-integrability-condition}
        \exists \ a\in I\ : \int_a^{\sup I}m(v)\,dv=\infty,
     \end{equation}
     Then $\left( M,g\right)$ is future $C^0$-inextendible. 
\end{theorem}
\begin{remark}[Interpretation of \Cref{thm:main thm3}]
\label{nonintcond-rem}
Under the assumptions $m>0$ and $m'\geq 0$, the non-integrability condition \eqref{non-integrability-condition} has different implications depending on whether $\sup I$ is finite or infinite.

Suppose first that $\sup I=\infty$. Then, for every $a\in I$ and every $R>a$, we have
$$
\int_a^R m(v)\,dv\geq m(a)(R-a),
$$
and therefore $\int_a^\infty m(v)\,dv=\infty$. Thus, in this case, condition \eqref{non-integrability-condition} is automatic and does not imply that the mass becomes unbounded. Since $m$ is nondecreasing, it may either converge to a finite limit or diverge to infinity. For example, a constant positive mass (as in the Schwarzschild spacetime with $I=\mathbb R$) satisfies the non-integrability condition, but trivially does not blow up as $v\to \infty$. On the other hand, a mass function of the form $m(v)=m_0e^v,\ m_0\in(0,\infty)$ is strictly increasing but diverges as $v\to\infty$.

Now suppose that $\sup I<\infty$. Since $m$ is nondecreasing, the limit $\lim_{v\to \sup I}m(v)$ exists in $(0,\infty]$. If this limit were finite, then $m$ would be bounded on $(a,\sup I)$ and $\int_a^{\sup I} m(v)\,dv$ would be finite. Hence condition \eqref{non-integrability-condition} forces
$$
\lim_{v\to \sup I}m(v)=\infty.
$$
In fact, the non-integrability condition would then require the mass to diverge sufficiently rapidly near the finite endpoint. Divergence of $m$ would not be enough: for example, $m(v)=(\sup I-v)^{-1/2}$ diverges as $v\to \sup I$ but is integrable there.

 Hence \Cref{thm:main thm3} may be interpreted as follows: ingoing Vaidya spacetimes with positive mass satisfying the weak energy condition are future $C^0$-inextendible whenever the infalling mass remains positive for arbitrarily long (advanced) times or grows without bound sufficiently rapidly in the case of a finite-time collapse.
 
\end{remark}

Under time reversal, \Cref{thm:main thm3} is equivalent to the following statement for outgoing Vaidya
spacetimes: \emph{Let $J=-I$, $\bar m(u)=m(-u)$, and $c=-a$. If $\bar m'(u)\le0$ and
$$\int_{\inf J}^{c}\bar m(u)\,du=\infty,$$
then the corresponding outgoing Vaidya spacetime is past $C^0$-inextendible.}

The main feature of this work is the extension of Sbierski’s argument to a setting where the translation isometry $(t,r,\omega)\mapsto(t+c,r,\omega)$ of the Schwarzschild interior is no longer available. In Schwarzschild spacetime, this translation preserves the metric and allows timelike curves to be moved while retaining their timelikeness. Sbierski explicitly uses this isometry in his proof of $C^0$-inextendibility of maximal analytic Schwarzschild spacetime. For ingoing Vaidya spacetimes with nonconstant mass function, such translations are no longer isometries. Similar difficulties show up at various stages of the proof due to the dynamical nature of the Vaidya spacetimes.

\color{black}

\subsection{Outline of the paper}

The paper is organised as follows. In \Cref{sec-2} we establish relevant properties of ingoing Vaidya spacetimes required to prove \Cref{thm:main thm3}.
In \Cref{sec-3} we prove \Cref{thm:main thm3} using Sbierski's tools. 
In \Cref{sec-4}, we highlight the additional technical arguments over Sbierski's original proof of $C^0$-inextendibility of maximal analytic Schwarzschild spacetime, needed to prove our main result. 
\color{black}
\section{The geometry of Vaidya spacetimes}\label{sec-2}

In $(M,g)$ as given in \Cref{Vaidya-spacetime}, the hypersurface $\mathcal{H^+}\coloneqq\{(v,r,\omega)\in M: r=2m(v)\}$ is called the \emph{apparent horizon} of the ingoing Vaidya spacetime, and has causal character as follows.

\begin{lemma}[Causal character of apparent horizon]\label{apparent-horizon}
    The apparent horizon of the ingoing Vaidya spacetime is spacelike wherever $m'(v)>0$, null wherever $m'(v)=0$ and timelike wherever $m'(v)<0$.
\end{lemma}
\begin{proof}
Let $F(v,r,\omega)\coloneqq r-2m(v)$. Then $\mathcal{H^+}=F^{-1}(0)$, and $dF=dr-2m'(v)\ dv$ is nowhere vanishing. Hence $\mathcal H^+$ is a $C^2$ embedded hypersurface of $M$, and $dF$ is a normal covector field to $\mathcal H^+$.
Using the inverse metric
$$
(g^{ab})_{a,b\in\{v,r\}}
=
\begin{pmatrix}
0 & 1\\
1 & \mathcal F
\end{pmatrix},
$$
we obtain
$$
g^{-1}(dF,dF)
=
-4m'(v)+1-\frac{2m(v)}{r}.
$$
On $\mathcal H^+$, where $r=2m(v)$, this becomes
$
\left.g^{-1}(dF,dF)\right|_{\mathcal H^+}
=
-4m'(v).
$
Hence the normal is timelike, null, or spacelike according as
$m'(v)>0$, $m'(v)=0$, or $m'(v)<0$. Therefore $\mathcal H^+$ is
respectively spacelike, null, or timelike.

    
\end{proof}

The apparent horizon separates the Vaidya spacetime into two regions,
\begin{equation*}
    M_\mathrm{ext} \coloneqq \{(v,r,\omega)\in M:r>2m(v)\} \quad \mathrm{and}\quad M_\mathrm{int}\coloneqq \{(v,r,\omega)\in M:r<2m(v)\}.
\end{equation*}

\begin{figure}
    \centering
    \includegraphics[width=\linewidth]{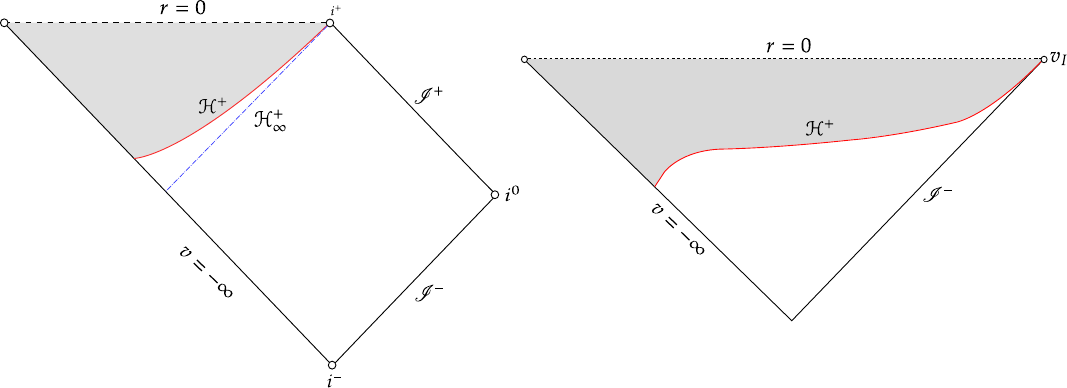}
    \caption{Examples of Penrose diagrams for ingoing Vaidya spacetimes
    admitting (left) and not admitting (right) a future null infinity.
    In the right diagram, $v_I$ represents the limit $v\to\sup I\leq\infty$.
    In both diagrams, the shaded region is $M_\mathrm{int}$ and the
    unshaded region is $M_\mathrm{ext}$.}
    \label{fig:vaidya-penrose}
\end{figure}

We call $M_{\mathrm{ext}}$ the \emph{exterior} and $M_{\mathrm{int}}$ the \emph{interior} of the Vaidya spacetime. See the left panel of \Cref{fig:vaidya-penrose} for a possible ``Schwarzschild-like" Penrose diagram of an ingoing Vaidya spacetime that admits a future null-infinity.

For spacetimes admitting a conformal completion $\bar M$ with
future null infinity $\mathscr I^+$ (see Definition 3.1.1 in \cite{chrusciel2020geometry}), the \emph{future event horizon} is defined by
$$
\mathcal H_\infty^+
:= \partial J^-(\mathscr I^+, \bar M)\cap M.
$$


Note that not all ingoing Vaidya spacetimes included in \Cref{thm:main thm3} have Penrose diagrams that look like the left panel of \Cref{fig:vaidya-penrose}. In particular, an arbitrary ingoing Vaidya spacetime need not even admit a future null infinity (see the right panel of \Cref{fig:vaidya-penrose}). See \cite{Berezin:2016ubu} for a discussion of how the behaviour of the mass function can influence the asymptotic structure of Vaidya spacetimes.

\begin{lemma}[Time functions in Vaidya interior]
\label{prop:timefunction}
    Any future-directed timelike curve in $M_\mathrm{int}$ can be parameterised by the radial coordinate $r$ or the advanced time coordinate $v$.
\end{lemma}
\begin{proof}
    Let $\gamma(s)=\bigl(\gamma_v(s),\gamma_r(s),\gamma_\omega(s)\bigr)$
be a future-directed timelike curve in $M_{\mathrm{int}}$. Since
$
dv^\sharp=\partial_r,
$
and $\partial_r$ is past-directed null, we have
$
\dot{\gamma}_v
=
dv(\dot{\gamma})
=
g(\partial_r,\dot{\gamma})
>0.
$
Hence $v$ is strictly increasing along $\gamma$, and therefore may be used as a parameter. Again, since $\gamma$ is timelike, we have,
    \begin{equation*}
         0>g(\dot\gamma(s),\dot\gamma(s)) =-\mathcal F(v,r) \dot\gamma_v(s)^2 +2\dot\gamma_v(s)\dot\gamma_r(s)
         +\gamma_r(s)^2 \| \dot\gamma_\omega (s)\|^2_{\mathbb S^2}.
    \end{equation*}
    Since we are working in $M_{\operatorname{int}}$, the first term on the RHS of the above inequality is positive. Also, the third term on the RHS  is non-negative. Since $\dot\gamma_v(s)>0$, we find that $\dot\gamma_r(s)<0$. Hence $r$ is strictly decreasing along $\gamma$ and may also be used as a parameter.
    
\end{proof}

\begin{remark}
\label{rem:vparaminM}
Since $dv^\sharp=\partial_r$ throughout $M$ and $\partial_r$ is past-directed null, every future-directed timelike curve $\gamma$ satisfies $dv(\dot\gamma)=g(\partial_r,\dot\gamma)>0$. A future-directed radial null curve has either the ingoing direction when its tangent is proportional to $-\partial_r$, for which $dv=0$, or the outgoing direction, for which $dv(\dot\gamma)>0$. Hence $v$ may be used as a parameter along every future-directed timelike curve and every future-directed outgoing radial null curve in the whole spacetime $(M,g)$ as well. By contrast, the conclusion $\dot\gamma_r<0$ uses essentially that $\mathcal F<0$ on $M_{\mathrm{int}}$. In $M_{\mathrm{ext}}$, where $\mathcal F>0$, the radial coordinate need not be monotone and therefore cannot generally be used as a parameter.
\end{remark}

The following lemma is a version of Lemma 3.4 in \cite{Sbierski_2018} adapted to the case of ingoing Vaidya spacetimes.

\color{black}

\begin{lemma}[Convergence of $v$ and $\omega$]
\label{lemma:vel-bounds}
    Let $(M,g)$ be an ingoing Vaidya spacetime satisfying $m'(v)\geq 0$, and let $p=(v_p,r_p,\omega_p)\in M_{\mathrm{int}}$. For every $\varepsilon>0$, there exists $0<\tilde r_p<r_p$ such that any future-directed timelike curve $\sigma:[-r_p,0)\to M_\mathrm{int}$ given by $\sigma(s) = (\sigma_v(s),-s,\sigma_\omega(s))$, with $\sigma(-r_p)=p$, satisfies 
\begin{equation}
    \label{velocity-bounds}
    |\sigma_v(s) - \sigma_v(s')|<\varepsilon, \qquad d_{\mathbb{S}^2}(\sigma_{\omega}(s),\sigma_\omega(s'))<\varepsilon, \qquad \forall ~-\tilde r_p \leq s,s'<0.
\end{equation}
In particular, $\sigma_v(s)$ and $\sigma_\omega(s)$ converge as $s\to 0$.
\end{lemma}
\begin{proof}
    Let $\sigma$ be a curve as in the statement, and set $c_p \coloneqq  2m(v_p)-r_p>0$. Timelikeness of $\sigma$ gives
    \begin{equation*}
        0>\left(\frac{2m(\sigma_v(s))}{-s}-1\right) \dot \sigma_v(s)^2 -2\dot\sigma_v(s)+ (-s)^2 \|\dot\sigma_\omega(s)\|_{\mathbb{S}^2}^2.
    \end{equation*}
Since $\dot\sigma_v(s)>0$, we have $\sigma_v(s)\geq v_p$, and $m'(v)\geq 0$ implies $m(\sigma_v(s))\geq m(v_p)>0$ for all $s\in [-r_p,0)$. Since $-s\leq r_p$ we obtain $2m(\sigma_v(s))-(-s)\geq2m(v_p)-r_p=c_p$. It follows from the timelikeness of $\sigma$ that
\begin{equation}
\label{v-convergence}
    |\dot\sigma_v(s)|<\frac{2(-s)}{2m(\sigma_v(s))-(-s)}\leq \frac{2(-s)}{c_p},
\end{equation}
and
\begin{equation}
    \label{omega-convergence}
    \|\dot\sigma_\omega(s)\|_{\mathbb{S}^2}  \leq \frac{2}{\sqrt{c_p(-s)}}.
\end{equation}
for all $s\in[-r_p,0)$. Both upper bounds are integrable on $(-r_p,0)$ and depend only on the fixed past endpoint $p$ and not the particular curve $\sigma$. Hence for every $\varepsilon>0$, one can choose $0<\tilde r_p<r_p$ such that integration over $(-\tilde r_p,0)$ gives the bounds in \eqref{velocity-bounds}. This completes the proof.

\end{proof} 

Next we state a result by Sbierski about the completeness of the exterior of spherically symmetric black hole spacetimes, which will be used to establish a geodesic completeness dichotomy for ingoing Vaidya spacetimes.

\begin{theorem}[Theorem A.2. and Remark A.4 in \cite{Sbierski:2020toa}]\label{ThmA2}
   Consider a general spherically symmetric spacetime $(M,g)$, where $M=\mathcal{Q}\times\mathbb{S}^2$ with $\mathcal{Q} \coloneqq [0,\infty)\times (0,\infty)$, and $g$ is given in the standard coordinates $(v,r,\omega)$ by,
   $$g=-\mathcal F(v,r)\ dv^2 + \frac{1}{2}h(v,r) (dv\otimes dr+ dr\otimes dv) + r^2 g_{\mathbb{S}^2}.$$ 
   Here $g_{\mathbb{S}^2}$ is the round metric on $\mathbb{S}^2$, and  $\mathcal F,h:\mathcal{Q} \to \mathbb{R}$ are $C^2$-functions. Declare a time orientation on $(M,g)$ by requiring $-\partial_r$ to be future-directed null. Let $r_{\mathcal{H}^+}:[0,\infty) \to (c_{\mathcal{H^+}},\infty)$ be continuous and uniformly positive, i.e., $c_{\mathcal{H^+}}>0$. Further define a curve $\sigma_{\mathcal{H}^+}:[0,\infty) \to \mathcal{Q}$ by $\sigma_{\mathcal{H^{+}}}\coloneqq (v,r_{\mathcal{H^+}}(v))$. The curve $\sigma_{\mathcal{H}^+}$ separates $M$ into two distinct regions, $$M_\mathrm{ext} \coloneqq \{(v,r,\omega)\in M:r>r_{\mathcal{H^+}}(v)\} \quad \mathrm{and}\quad M_\mathrm{int}\coloneqq \{(v,r,\omega)\in M:r<r_{\mathcal{H}^+}(v)\}.$$ Let $\gamma:[0,b)\to M$ be an affinely parameterized future-directed future-inextendible timelike geodesic with $b\in(0,\infty]$ and $\gamma(0) \in M_\mathrm{ext}$. Assume moreover that $\mathcal F$ and $h$ satisfy on $M_\mathrm{ext} \cap I^{+}(\gamma(0), M)$,
   \begin{equation*}
       |\mathcal F|\leq C_\mathcal F, \quad 0<c_h\leq h\leq C_h, \quad \partial_r h\geq 0, \quad \mathcal F\partial_rh -h\partial_r \mathcal F\leq0,
   \end{equation*}
   for some constants $C_\mathcal F, c_h,C_h$ and $C$. Then $\gamma$ is future complete, or there exists $\lambda_0\in[0,b)$ with $\gamma(\lambda_0) \in \mathrm{Im}(\sigma_{\mathcal{H}^+})$. 
\end{theorem}

\begin{remark}[Validity of \Cref{ThmA2} under translation of the $v$-coordinate]\label{rem-val-thmA2}
\Cref{ThmA2} remains valid if $Q=[0,\infty)\times(0,\infty)$ is replaced by $Q_a=[a,\infty)\times(0,\infty)$ for any $a\in\mathbb{R}$. Indeed, by defining a new translated coordinate $u\coloneqq v-a$ one has $Q_a=[0,\infty)_u\times(0,\infty)$ and $dv=du$. Defining $\tilde {\mathcal F}(u,r)\coloneqq \mathcal F(u+a,r),\ \tilde h(u,r)\coloneqq h(u+a,r),$ and $\tilde r_{\mathcal H^+}(u)\coloneqq r_{\mathcal H^+}(u+a)$, preserves all the hypotheses of \Cref{ThmA2}. Therefore, the proof of \Cref{ThmA2} as presented in \cite{Sbierski:2020toa} applies without change. 
\end{remark}

\begin{theorem}[Timelike geodesic dichotomy for ingoing Vaidya spacetime]\label{timelike-incompletness}
Let $(M,g)$ be an ingoing Vaidya spacetime as in \Cref{Vaidya-spacetime}, satisfying the weak energy condition $m'(v)\geq 0$ and the non-integrability condition \eqref{non-integrability-condition}. A future-directed future-inextendible timelike geodesic $\gamma:[0,s_+)\to M$ is future incomplete if and only if $r\circ\gamma(s)\to 0$ as $s\to s_+$. 
\end{theorem}
\begin{proof} We first consider the case $\sup I=\infty$. The proof in this case proceeds in seven steps:

\smallskip
\underline{\textit{Step 1}: Affine reparametrization.}
Since $\gamma$ is timelike we can consider its parametrization by proper time $\bar\gamma:[0,\lambda_+)\to M$, $\bar\gamma(\lambda)=(\bar\gamma_v(\lambda),\bar\gamma_r(\lambda),\bar\gamma_\omega(\lambda))$ ($\lambda\in [0,\lambda_+)$).
It follows that:
\begin{enumerate}
\item $\gamma$ is future incomplete if and only if $\lambda_+<\infty$;
\item $\gamma_r(s)\to 0$ as $s\to s_+$ if and only if $\bar\gamma_r(\lambda)\to 0$ as $\lambda\to\lambda_+$ since the reparametrization is strictly increasing.
\end{enumerate}
Hence it suffices to work with $\bar\gamma$ and the affine parameter $\lambda$.\\
\smallskip
\underline{\textit{Step 2}: Christoffel symbols for the ingoing Vaidya spacetime.}
The non-zero Christoffel symbols for the ingoing Vaidya metric (see \cite{Coudray:2021pjr}) are,
\begin{align*}
&\Gamma^{v}_{vv}=\frac{m(v)}{r^{2}},\qquad
\Gamma^{v}_{\theta\theta}=-r,\qquad
\Gamma^{v}_{\varphi\varphi}=-r\sin^{2}\theta,\\
&\Gamma^{r}_{vv}=\frac{m'(v)}{r}+\mathcal F(v,r)\frac{m(v)}{r^{2}},\qquad
\Gamma^{r}_{vr}=\Gamma^{r}{}_{rv}=-\frac{m(v)}{r^{2}},\\
&\Gamma^{r}_{\theta\theta}=-\mathcal F(v,r)\,r,\qquad
\Gamma^{r}_{\varphi\varphi}=-\mathcal F(v,r)\,r\sin^{2}\theta,\\
&\Gamma^{\theta}_{r\theta}=\Gamma^{\theta}_{\theta r}=\frac{1}{r},\qquad
\Gamma^{\theta}_{\varphi\varphi}=-\sin\theta\cos\theta,\\
&\Gamma^{\varphi}_{r\varphi}=\Gamma^{\varphi}_{\varphi r}=\frac{1}{r},\qquad
\Gamma^{\varphi}_{\theta\varphi}=\Gamma^{\varphi}_{\varphi\theta}=\cot\theta.
\end{align*}
here $\mathcal F(v,r)$ is the function defined in \eqref{vaidya-original-metric}. All other $\Gamma^\mu_{\alpha\beta}$ vanish.

\smallskip
\underline{\textit{Step 3}: $\mathrm{SO}(3)$ Killing vector fields.}
With $\omega\equiv(\theta,\varphi)$ the standard spherical coordinates on $\mathbb{S}^{2}$, we have $g_{\mathbb{S}^{2}}=d\theta\otimes d\theta+\sin^{2}\theta\ d\varphi\otimes d\varphi$.
By spherical symmetry, the ingoing Vaidya spacetime admits the standard rotational Killing fields on $\mathbb{S}^{2}$ (see section 3.8 of \cite{Carroll:2004st}),
$$
K_{1}=-\sin\varphi\ \partial_{\theta}-\cot\theta\cos\varphi\ \partial_{\varphi}, \quad
K_{2}= \cos\varphi\ \partial_{\theta}-\cot\theta\sin\varphi\ \partial_{\varphi}, \quad
K_{3}= \partial_{\varphi}.
$$
Since $\bar\gamma$ is affinely parametrized and $K_i$ are Killing fields for $i\in\{1,2,3\}$, the quantities
$$
J_i:=g(K_i(\bar\gamma(\lambda)),\dot{\bar\gamma}(\lambda)),\quad i\in\{1,2,3\}
$$
are constant in $\lambda$, i.e., 
$$
\frac{d}{d\lambda}g(K_i(\bar\gamma(\lambda)),\dot{\bar\gamma}(\lambda))
= g(\nabla_{\dot{\bar\gamma}(\lambda)}K_i(\bar\gamma(\lambda)),\dot{\bar\gamma}(\lambda)) + g(K_i(\bar\gamma(\lambda)),\nabla_{\dot{\bar\gamma}(\lambda)}\dot{\bar\gamma}(\lambda))
=0+0=0.
$$
We write $\bar\gamma_\omega(\lambda)=(\bar\gamma_\theta(\lambda),\bar\gamma_\varphi(\lambda))$. Along $\bar\gamma$ we have, $g_{\theta\theta}\circ\gamma=\bar\gamma_r^{2}$, $g_{\varphi\varphi}\circ\gamma=\bar\gamma_r^{2}\sin^{2}\bar\gamma_\theta$,
$g_{\theta\varphi}\circ\gamma=0$. Hence along $\bar\gamma$
\begin{align*}
J_{1}&=\bar\gamma_r(\lambda)^{2}\Big(-\sin\bar\gamma_\varphi(\lambda)\ \dot{\bar\gamma}_\theta(\lambda)
-\sin\bar\gamma_\theta(\lambda)\cos\bar\gamma_\theta(\lambda)\cos\bar\gamma_\varphi(\lambda)\ \dot{\bar\gamma}_\varphi(\lambda)\Big),\\
J_{2}&=\bar\gamma_r(\lambda)^{2}\Big( \cos\bar\gamma_\varphi(\lambda)\ \dot{\bar\gamma}_\theta(\lambda)
-\sin\bar\gamma_\theta(\lambda)\cos\bar\gamma_\theta(\lambda)\sin\bar\gamma_\varphi(\lambda)\ \dot{\bar\gamma}_\varphi(\lambda)\Big),\\
J_{3}&=\bar\gamma_r(\lambda)^{2}\sin^{2}\bar\gamma_\theta(\lambda)\ \dot{\bar\gamma}_\varphi(\lambda).
\end{align*}
Define the \emph{angular momentum} $L\geq 0$ by $L^{2}:=J_{1}^{2}+J_{2}^{2}+J_{3}^{2}.$ Then $L$ is constant in $\lambda$. One can simplify further to obtain,
\begin{equation}
\label{L-identity}
L^{2} =\bar\gamma_r(\lambda)^{4}\Big(\dot{\bar\gamma}_\theta^{2}(\lambda) +\sin^{2}\bar\gamma_\theta(\lambda)\ \dot{\bar\gamma}_\varphi^{2}(\lambda)\Big) =\bar\gamma_r(\lambda)^{4}\ \|\dot{\bar\gamma}_\omega(\lambda)\|_{\mathbb{S}^{2}}^{2}.
\end{equation}
In particular we have,
\begin{equation}
\label{omega-dot-bound-formula}
\|\dot{\bar\gamma}_\omega(\lambda)\|_{\mathbb{S}^{2}}^{2}=\frac{L^{2}}{\bar\gamma_r(\lambda)^{4}}.
\end{equation}

\smallskip
\underline{\textit{Step 4}: Estimate for $\dot{\bar\gamma}_r$ on $M_{\mathrm{int}}$.}
From the proper-time normalization of Step 1, we have
\begin{equation} \label{normalization-raw}
\begin{aligned}
-1&=g(\dot{\bar\gamma}(\lambda),\dot{\bar\gamma}(\lambda))\\
&=-\mathcal F(\bar\gamma_v(\lambda),\bar\gamma_r(\lambda))\,\dot{\bar\gamma}_v(\lambda)^{2}
+2\dot{\bar\gamma}_v(\lambda) \dot{\bar\gamma}_r(\lambda)
+\bar\gamma_r(\lambda)^{2}\|\dot{\bar\gamma}_\omega(\lambda)\|_{\mathbb{S}^{2}}^{2},
\end{aligned}
\end{equation}
 where $\mathcal F$ is defined in  \eqref{vaidya-original-metric}.  Using \eqref{omega-dot-bound-formula}, we have
\begin{equation}
\label{normalization}
-1=-\mathcal F(\bar\gamma_v(\lambda),\bar\gamma_r(\lambda))\ \dot{\bar\gamma}_v(\lambda)^{2}
+2\dot{\bar\gamma}_v(\lambda)\dot{\bar\gamma}_r(\lambda)
+\frac{L^{2}}{\bar\gamma_r(\lambda)^{2}}\,,
\end{equation}
which can be treated as a quadratic equation in the unknown $\dot{\bar\gamma}_v(\lambda)$. On $M_{\mathrm{int}}$ we have $\mathcal F<0$. The discriminant of  \eqref{normalization} is
$$
\Delta=(2\dot{\bar\gamma}_r(\lambda))^{2}+4\mathcal F(\bar\gamma_v(\lambda),\bar\gamma_r(\lambda))\left(1+\frac{L^{2}}{\bar\gamma_r^{2}(\lambda)}\right).
$$
Since $\dot{\bar\gamma}_v(\lambda)$ is (in particular) real, necessarily $\Delta\geq 0$. Hence on $M_{\mathrm{int}}$ we have the estimate,
\begin{equation}
\label{discriminant-ineq}
(\dot{\bar\gamma}_r(\lambda))^{2}\geq -\mathcal F(\bar\gamma_v(\lambda),\bar\gamma_r(\lambda))\left(1+\frac{L^{2}}{\bar\gamma_r(\lambda)^{2}}\right)\geq -\mathcal F(\bar\gamma_v(\lambda),\bar\gamma_r(\lambda)).
\end{equation}

\smallskip
\underline{\textit{Step 5}: If $\bar\gamma$ starts in $M_{\mathrm{ext}}$ and is incomplete, then $\bar\gamma$ enters $M_{\mathrm{int}}$.} Suppose $\bar\gamma(0)\in M_{\mathrm{ext}}$ and choose $a\in I$ with $a<\bar\gamma_v(0)$. Since $\dot{\bar\gamma}_v>0$ along every future-directed timelike curve, $\bar\gamma$ remains in (the topological) interior of the region
$$
M_a \coloneqq 
\bigl\{(v,r,\omega)\in M:v\geq a\bigr\}=[a,\infty)\times(0,\infty)\times\mathbb S^2\subseteq M.$$
On $M_a$, one has $2m(v)\geq 2m(a)>0$. Applying \Cref{ThmA2} to $(M_a,g|_{M_a})$ \footnote{Using \Cref{rem-val-thmA2} it is easy to check that the hypotheses of \Cref{ThmA2} hold for $(M_a,g|_{M_a})$}, one obtains that $\bar\gamma$ is future complete, or $\bar\gamma$ meets the apparent horizon $\mathcal{H^+}=\{r=2m(v)\}$ at some parameter value. Equivalently if $\bar\gamma$ does not meet $\mathcal{H}^+$, then it is future complete. Taking the contrapositive yields,
\begin{equation}
\label{thm39-contrapositive}
\bar\gamma(0)\in M_{\mathrm{ext}}\ \text{ and }\ \bar\gamma\ \text{future incomplete}
\quad\Longrightarrow\quad \exists \lambda_1 > 0 :\ \bar\gamma_r(\lambda_1)=2m(\bar\gamma_v(\lambda_1)).
\end{equation}

Set $F(v,r,\omega)\coloneqq r-2m(v)$. By \Cref{apparent-horizon}, the normal $\nabla F|_{\mathcal H^+}=\partial_v-2m'(v)\partial_r$ is causal, and it is future-directed since $g(\nabla F,-\partial_r)=-1$. Thus $(F\circ\bar\gamma)'(\lambda)=g(\nabla F\circ\bar\gamma(\lambda),\dot{\bar\gamma}(\lambda))<0$ at every intersection with $\mathcal H^+$. Consequently, $\bar\gamma$ enters $M_{\mathrm{int}}$ immediately after such an intersection and cannot subsequently leave it. Together with \eqref{thm39-contrapositive}, this shows that every future-incomplete $\bar\gamma$ eventually remains in $M_{\mathrm{int}}$. Hence, for the implication ``future incomplete $\implies\bar\gamma_r(\lambda)\to0$'', it suffices to work on an interval
\begin{equation}
\label{terminal-int-interval}
    [\lambda_*,\lambda_+)\subset[0,\lambda_+)\quad \text{such that} \quad  \bar\gamma(\lambda)\in M_{\mathrm{int}} \quad \text{for all} \quad \lambda\in[\lambda_\ast,\lambda_+).
\end{equation}

\smallskip
\underline{\textit{Step 6}: $\bar\gamma_r(\lambda)\to 0 \ \Rightarrow\ $ future incompleteness.}
Assume $\bar\gamma_r(\lambda)\to 0$ as $\lambda\to\lambda_+$, and let $m_0\coloneqq m(\bar\gamma_v(0))>0$. Since $\dot{\bar\gamma}_v>0$ and $m'(v)\geq 0$, we have 
\begin{equation}
    \label{lower-bd}
    m(\bar\gamma_v(\lambda))\geq m_0 \qquad \text{for every $\lambda\in[0,\lambda_+)$}.
\end{equation}
Now for $\lambda$ close to $\lambda_+$ we have
$\bar\gamma(\lambda)\in M_{\mathrm{int}}$ and so we may use \eqref{discriminant-ineq} and \eqref{lower-bd} along with $\dot{\bar\gamma}_r<0$ (see  \Cref{prop:timefunction}) to get
$$
-\dot{\bar\gamma}_r (\lambda)\geq \sqrt{-\mathcal F(\bar\gamma_v(\lambda),\bar\gamma_r(\lambda))} =\sqrt{\frac{2m(\bar\gamma_v(\lambda))}{\bar\gamma_r(\lambda)}-1} \geq \sqrt{\frac{2m_0}{\bar\gamma_r(\lambda)}-1}.
$$
 For $\bar\gamma_r(\lambda)\leq m_0$ we have $(2m_0/\bar\gamma_r(\lambda))-1\geq m_0/\bar\gamma_r(\lambda)$, and hence for $\lambda$ sufficiently close to $\lambda_+$,
$$ -\dot{\bar\gamma}_r(\lambda)\geq \sqrt{\frac{m_0}{\bar\gamma_r(\lambda)}}.$$ 
Integrating from $\lambda$ to $\lambda_+$ and using that $\bar\gamma_r(\lambda)\to 0$,
$$
\lambda_+-\lambda
=\int_{\lambda}^{\lambda_+} d\sigma
\leq -\int_{\bar\gamma_r(\lambda)}^{0}\sqrt{\frac{r}{m_0}} dr
=\frac{2}{3}\frac{\bar\gamma_r(\lambda)^{3/2}}{\sqrt{m_0}}<\infty.
$$
Hence $\lambda_+<\infty$, i.e.\ $\bar\gamma$ is future incomplete.

\smallskip
\underline{\textit{Step 7}: Future incompleteness $\Rightarrow \bar\gamma_r(\lambda)\to 0$.}
Assume $\bar\gamma$ is future incomplete, so $\lambda_+<\infty$. By \eqref{terminal-int-interval}, fix $\lambda_\ast<\lambda_+$ such that $\bar\gamma(\lambda)\in M_{\mathrm{int}}$ for all $\lambda\in[\lambda_\ast,\lambda_+)$. As $\bar\gamma_r$ is decreasing and positive on $[\lambda_*,\lambda_+)$, its limit as $\lambda\to \lambda_+$ exists in $[0,\infty)$. Suppose, for contradiction, that
$$r_*:=\lim_{\lambda\to\lambda_+}\bar\gamma_r(\lambda)>0.$$
Choose $\lambda_1\in[\lambda_\ast,\lambda_+)$ such that for all $\lambda\in[\lambda_1,\lambda_+)$,
\begin{equation}
\label{r-bounds}
\bar\gamma_r(\lambda)\in[r_-,r_+],\qquad r_-:=r_*,\quad r_+:=\bar\gamma_r(\lambda_1).
\end{equation}
On $M_\mathrm{int}$ we have $\dot{\bar\gamma}_v(\lambda)>0$ (see  \Cref{prop:timefunction}). The $v$-geodesic equation (computed from the Christoffel symbols from Step 2) reads,
\begin{equation}
\label{v-geodesic-L}
\ddot{\bar\gamma}_v(\lambda) =-\frac{m(\bar\gamma_v(\lambda))}{\bar\gamma_r(\lambda)^{2}}(\dot{\bar\gamma}_v(\lambda))^{2}+\frac{L^{2}}{\bar\gamma_r(\lambda)^{3}}\,.
\end{equation}
Let $m_1\coloneqq m(\bar\gamma_v(\lambda_1))>0$, then $m(\bar\gamma_v(\lambda))\geq m_1$ for every $\lambda \in [\lambda_1,\lambda_+)$. Using this and $\bar\gamma_r(\lambda)\in [r_-,r_+]$ on $[\lambda_1,\lambda_+)$ in \eqref{v-geodesic-L}, we have
$$
\ddot{\bar\gamma}_v(\lambda)\leq -\frac{m_1}{r_+^{2}}\dot{\bar\gamma}_v(\lambda)^{2}+\frac{L^{2}}{r_-^{3}}.
$$
Set $a:=m_1/r_+^2>0$ and $b:=L^2/r_-^3\ge 0$. Then
\begin{equation}
    \label{diff-ineq}
\ddot{\bar\gamma}_v(\lambda) \le -a\dot{\bar\gamma}_v(\lambda)^{2}+b = -a\left(\dot{\bar\gamma}_v(\lambda)^{2}-\frac{b}{a}\right).
\end{equation}
We claim that $\dot{\bar\gamma}_v(\lambda)\le \max\{\dot{\bar\gamma}_v(\lambda_1),\sqrt{b/a}\}$ for all $\lambda\in[\lambda_1,\lambda_+)$. Assume for contradiction that there exists $\lambda_2\in(\lambda_1,\lambda_+)$ with
$\dot{\bar\gamma}_v(\lambda_2)>\max\{\dot{\bar\gamma}_v(\lambda_1),\sqrt{b/a}\}=: C_v$, and let
$$
\Lambda\coloneqq \Big\{\lambda\in[\lambda_1,\lambda_2]:\dot{\bar\gamma}_v(\lambda)\leq C_v\Big\}, \quad\lambda_0:=\sup \Lambda.
$$

Since $\dot{\bar\gamma}_v$ is continuous, $\Lambda$ is closed in the compact interval $[\lambda_1,\lambda_2]$, and hence is compact. Moreover, $\lambda_1\in\Lambda$, so $\Lambda\neq\varnothing$. Therefore $\lambda_0=\sup\Lambda=\max\Lambda\in\Lambda$. Since $\lambda_2\notin\Lambda$, we have $\lambda_0<\lambda_2$ and $\dot{\bar\gamma}_v(\lambda_0)\leq C_v.$ If $\dot{\bar\gamma}_v(\lambda_0)<C_v$, then, by continuity, there
exists $\varepsilon\in(0,\lambda_2-\lambda_0)$ such that $\dot{\bar\gamma}_v(\lambda)<C_v$ for all $\lambda\in[\lambda_0,\lambda_0+\varepsilon]$. Hence $\lambda_0+\varepsilon\in\Lambda$, contradicting $\lambda_0=\max\Lambda$. Hence $\dot{\bar\gamma}_v(\lambda_0)=C_v.$ Now by definition of $\lambda_0$, for every $\lambda\in(\lambda_0,\lambda_2]$ we have $\dot{\bar\gamma}_v(\lambda)>C_v\ge \sqrt{b/a}$. Using inequality \eqref{diff-ineq}, we obtain $\ddot{\bar\gamma}_v(\lambda)\le -a(\dot{\bar\gamma}_v(\lambda)^2-(b/a))<0$ for all $\lambda\in(\lambda_0,\lambda_2]$. So $\dot{\bar\gamma}_v$ is strictly decreasing on $(\lambda_0,\lambda_2]$. Hence $\dot{\bar\gamma}_v(\lambda_2)<\dot{\bar\gamma}_v(\lambda_0)=C_v$, contradicting $\dot{\bar\gamma}_v(\lambda_2)>C_v$. Hence we have
\begin{equation}
\label{vdot-upper}
0<\dot{\bar\gamma}_v(\lambda)\leq C_v\,,
\end{equation}
for all $\lambda\in[\lambda_1,\lambda_+)$.
It follows that $\bar\gamma_v$ is bounded on $[\lambda_1,\lambda_+)$, i.e.,
\begin{equation}
\label{v-range}
\forall\lambda\in[\lambda_1,\lambda_+) : \bar\gamma_v(\lambda)\in[v_-,v_+];
\quad
v_-:=\bar\gamma_v(\lambda_1), \quad v_+:=\bar\gamma_v(\lambda_1)+C_v(\lambda_+-\lambda_1).
\end{equation}
Define
$$
m_{+}:=\max_{[v_-,v_+]} m(v)<\infty.
$$
Using $m(\bar\gamma_v)\leq m_+$ and $\bar\gamma_r\geq r_-$ in \eqref{v-geodesic-L} gives
$$
\ddot{\bar\gamma}_v(\lambda)\geq -\frac{m_+}{r_-^{2}}\dot{\bar\gamma}_v(\lambda)^{2}.
$$
Let $A_0:=m_+/r_-^{2}$. Then $d(1/\dot{\bar\gamma}_v(\lambda))/d\lambda=-\ddot{\bar\gamma}_v(\lambda)/\dot{\bar\gamma}_v(\lambda)^{2}\leq A_0$, and hence for $\lambda\in[\lambda_1,\lambda_+)$,
$$
\frac{1}{\dot{\bar\gamma}_v(\lambda)}\leq \frac{1}{\dot{\bar\gamma}_v(\lambda_1)}+A_0(\lambda-\lambda_1)
\leq \frac{1}{\dot{\bar\gamma}_v(\lambda_1)}+A_0(\lambda_+-\lambda_1).
$$
It follows that,
\begin{equation}
\label{vdot-lower}
\dot{\bar\gamma}_v(\lambda)\geq c_v:=
\frac{1}{(\dot{\bar\gamma}_v(\lambda_1))^{-1}+A_0(\lambda_+-\lambda_1)}>0
\end{equation}
for all $\lambda\in[\lambda_1,\lambda_+)$. Now from \eqref{omega-dot-bound-formula} and $\bar\gamma_r(\lambda)\geq r_-$, we have
\begin{equation}
\label{omega-dot-bound}
\|\dot{\bar\gamma}_\omega(\lambda)\|_{\mathbb{S}^{2}}
\leq C_\omega:=\frac{L}{r_-^{2}}
\end{equation}
for all $\lambda\in[\lambda_1,\lambda_+)$. Solving \eqref{normalization} for $\dot{\bar\gamma}_r$ gives,
$$
\dot{\bar\gamma}_r(\lambda) =\frac{-1+\mathcal F(\bar\gamma_v(\lambda),\bar\gamma_r(\lambda))\ \dot{\bar\gamma}_v(\lambda)^{2}- (L^{2}/\bar\gamma_r(\lambda)^{2})}{2\dot{\bar\gamma}_v(\lambda)}.
$$
On $[\lambda_1,\lambda_+)$, we have $\bar\gamma_r(\lambda)\in[r_-,r_+]$ and $m(\bar\gamma_v(\lambda))\in[m_1,m_+]$, so
$$
|\mathcal F(\bar\gamma_v(\lambda),\bar\gamma_r(\lambda))|=\left|1-\frac{2m(\bar\gamma_v(\lambda))}{\bar\gamma_r(\lambda)}\right|
\leq 1+\frac{2m_+}{r_-}=:\mathcal F_+.
$$
Using $\dot{\bar\gamma}_v(\lambda)\in[c_v,C_v]$ from \eqref{vdot-upper}--\eqref{vdot-lower} and $\bar\gamma_r(\lambda)^{-2}\leq r_-^{-2}$ yields
\begin{equation}
\label{rdot-bound}
|\dot{\bar\gamma}_r(\lambda)| \leq  \frac{1+\mathcal F_+C_v^{2}+(L^{2}/r_-^{2})}{2c_v}=: C_r\,,
\end{equation}
for all $\lambda\in[\lambda_1,\lambda_+)$. Define
$$
D_{C_\omega}:=\{(\omega,\dot\omega)\in T\mathbb{S}^{2}:\ \|\dot\omega\|_{\mathbb{S}^{2}}\leq C_\omega\},
$$
and the compact set
$$
\mathcal K :=\Big\{(v,r,\omega,\dot v,\dot r,\dot \omega)\in TM:\  v\in[v_-,v_+],\ r\in[r_-,r_+],\ (\omega,\dot\omega)\in D_{C_\omega},\ \dot v\in[c_v,C_v],\ \dot r\in[-C_r,C_r]\Big\}.
$$
By \eqref{r-bounds} and \eqref{vdot-upper}--\eqref{rdot-bound}, we have for all $\lambda\in[\lambda_1,\lambda_+)$,
\begin{equation}
\label{phase-in-K}
(\bar\gamma(\lambda),\dot{\bar\gamma}(\lambda))\in\mathcal K.
\end{equation}

Affinely parametrized geodesics are the integral curves of a $C^1$ vector field $\mathcal G$ on $TM$:
\[
\frac{d}{d\lambda}(\bar\gamma(\lambda),\dot{\bar\gamma}(\lambda))
=\mathcal G(\bar\gamma(\lambda),\dot{\bar\gamma}(\lambda)),
\qquad 
\mathcal G(x,u)\coloneqq(u,-\Gamma(x)(u,u)),
\]
where $\Gamma$ denotes the \emph{Christoffel map}, i.e., in coordinates $x^{\rho}$ we have $(\Gamma(x)(u,u))^{\rho} = \Gamma^{\rho}_{\mu\nu}(x)u^\mu u^\nu$. Since $\bar\gamma_r(\lambda)\geq r_->0$ on $\mathcal K$, all Christoffel symbols $\Gamma^{\rho}_{\mu\nu}$ (see Step 2) are $C^1$ on the base projection of $\mathcal K$, and hence $\mathcal G$ is locally Lipschitz on a neighbourhood of $\mathcal K$. By the standard continuation theorem for ODEs we have that a maximal integral curve can fail to extend beyond a finite time only if it leaves every compact subset before that time (see also \cite[Prop.\ 3.4, Rem.\ 3.5]{CS:08}). For the timelike geodesic $\bar\gamma$, this contradicts \eqref{phase-in-K}. Hence $(\bar\gamma(\lambda),\dot{\bar\gamma}(\lambda))$ extends beyond $\lambda_+$, contradicting future-inextendibility. Therefore the assumption $r_*>0$ is false. Hence $r_*=0$ and
this proves that future incompleteness implies $\bar\gamma_r\to 0$. This completes the proof in the case $\sup I=\infty$.

\medskip
It remains to consider the case $\sup I<\infty$. Let $\bar\gamma:[0,\lambda_+)\to M$ be the proper-time parametrization of $\gamma$ as in Step 1. The implication $\bar\gamma_r(\lambda)\to0\Rightarrow\lambda_+<\infty$ follows from Step 6, whose argument does not require $\sup I=\infty$. To prove the converse, suppose that $\lambda_+<\infty$.

Set $v_0\coloneqq\bar\gamma_v(0)$ and $r_0\coloneqq\bar\gamma_r(0)$. Since $\dot{\bar\gamma}_v>0$, timelikeness gives
\begin{equation}\label{eq:gam-TL}
\dot{\bar\gamma}_r(\lambda)
<\frac12\left(1-\frac{2m(\bar\gamma_v(\lambda))}{\bar\gamma_r(\lambda)}\right)\dot{\bar\gamma}_v(\lambda)
\le\frac12\dot{\bar\gamma}_v(\lambda).
\end{equation}
Consequently, for every $\lambda\in[0,\lambda_+)$,
$$
0<\bar\gamma_r(\lambda)
\le r_0+\frac{\bar\gamma_v(\lambda)-v_0}{2}
\le R\coloneqq r_0+\frac{\sup I-v_0}{2}.
$$
Multiplying the first differential inequality in \eqref{eq:gam-TL} by $2\bar\gamma_r(\lambda)$ therefore gives
$$
\frac{d}{d \lambda}({\bar\gamma}_r(\lambda)^2)=2\bar\gamma_r(\lambda)\dot{\bar\gamma}_r(\lambda)
<\bigl(R-2m(\bar\gamma_v(\lambda))\bigr)\dot{\bar\gamma}_v(\lambda).
$$
Integrating, we obtain for $0<\lambda<\lambda_+$,
$$
0<\bar\gamma_r(\lambda)^2
<r_0^2+R\bigl(\bar\gamma_v(\lambda)-v_0\bigr)
-2\int_{v_0}^{\bar\gamma_v(\lambda)}m(v)\,dv.
$$
Since $\bar\gamma_v$ is increasing and bounded above by $\sup I<\infty$, its limit $v_*\coloneqq\lim_{\lambda\to\lambda_+}\bar\gamma_v(\lambda)$ exists. If $v_*=\sup I$, the right-hand side tends to $-\infty$ by the non-integrability condition \eqref{non-integrability-condition}, which is a contradiction. Hence $v_*\in I$.

Moreover, $\bar\gamma_r-\bar\gamma_v/2$ is decreasing and bounded below by $-(\sup I)/2$. Since $\bar\gamma_v$ converges, it follows that $r_*\coloneqq\lim_{\lambda\to\lambda_+}\bar\gamma_r(\lambda)$ exists in $[0,\infty)$. Suppose that $r_*>0$. Choose $\lambda_1<\lambda_+$ so close to $\lambda_+$ that
$$
\bar\gamma_r(\lambda)\in[r_-,r_+],
\qquad
r_-\coloneqq r_*/2,\qquad r_+\coloneqq R,
\qquad \lambda\in[\lambda_1,\lambda_+).
$$
Set $v_-\coloneqq\bar\gamma_v(\lambda_1)$ and choose $v_+\in I$ with $v_*<v_+$. Then
$$
\bar\gamma_v(\lambda)\in[v_-,v_+]\Subset I,
\qquad
m_+\coloneqq\max_{[v_-,v_+]}m<\infty.
$$
With these bounds, the velocity estimates of Step 7 apply: they use only the geodesic equations, $\dot{\bar\gamma}_v>0$, positivity and monotonicity of $m$, and the bounds on $v$ and $r$. Thus there are positive constants $c_v,C_v,C_r,C_\omega<\infty$ such that
$$
c_v\le\dot{\bar\gamma}_v(\lambda)\le C_v,
\qquad
|\dot{\bar\gamma}_r(\lambda)|\le C_r,
\qquad
\|\dot{\bar\gamma}_\omega(\lambda)\|_{\mathbb S^2}\le C_\omega
$$
on $[\lambda_1,\lambda_+)$. Hence $(\bar\gamma,\dot{\bar\gamma})$ remains in a compact subset of $TM$. Since $\lambda_+<\infty$, the geodesic-flow continuation argument of Step 7 extends $\bar\gamma$ beyond $\lambda_+$, contradicting future-inextendibility. Therefore $r_*=0$, which completes the proof in the case $\sup I<\infty$ as well.
\color{black}

\end{proof}

\begin{proposition}[Global hyperbolicity]
\label{global-hyp}
    The ingoing Vaidya spacetimes as defined in \Cref{Vaidya-spacetime} are globally hyperbolic.
\end{proposition}
\begin{proof}
    Define $T\coloneqq v-r$. Since
    $$g^{-1}(dT,dT)=\mathcal F-2=-1-\frac{2m(v)}{r}<0, \qquad dT(-\partial_r)=1,$$
    hence $T$ is a temporal function on $M$ increasing towards the future. Hence by Theorem 4.100 of \cite{Minguzzi:2019mbe}, $(M,g)$ is stably causal, hence strongly causal.
It remains to prove the compactness of causal diamonds. So let $p=(v_p,r_p,\omega_p)\leq q=(v_q,r_q,\omega_q)\in M$ and assume $J(p,q)\coloneqq J^+(p,M)\cap J^-(q,M)\neq \emptyset$ (if it is empty then there is nothing to prove). If $p=q$ then causality gives $J(p,q)=\{p\}$. Hence assume also $p\neq q$. We first show that $J(p,q)$ is contained in a closed and bounded set (that we call $\mathcal K$). Then, we prove that  $J(p,q)$ is closed in $\mathcal K$.

Along any future-directed causal curve $\gamma:[0,1]\to M$ with $\gamma(0)=p$ and $\gamma(1)=q$, given by $\gamma(s)=(\gamma_v(s),\gamma_r(s),\gamma_\omega(s))$, we have
    $$\dot\gamma_v(s)=-g\big(\dot\gamma(s),-\partial_r(\gamma(s))\big)\geq 0,\qquad -\mathcal F(\gamma_v(s),\gamma_r(s))\dot\gamma_v^2+2\dot\gamma_v(s)\dot\gamma_r(s)+\gamma_r(s)^2\|\dot\gamma_{\omega}(s)\|_{\mathbb{S}^2}^2\leq 0.$$

 When $\dot\gamma_v(s)>0$, these imply 
    \begin{equation}
        \label{vdot-rdot}
       \dot\gamma_r(s)\leq \frac12\mathcal F(\gamma_v(s),\gamma_r(s))\dot\gamma_v(s)\leq \frac{1}{2}\dot\gamma_v(s).
    \end{equation}
     
        When $\dot\gamma_v(s)=0$, causality of $\gamma$ forces $\dot\gamma_\omega(s)=0$. 
        Additionally,  $-\partial_r$ is future-directed, and $\dot \gamma(s)=\dot\gamma_r(s)\partial_r=-\dot\gamma_r(s)\left(-\partial_r\right)$. Therefore, for $\gamma(s)$ to be future-directed, $-\dot\gamma_r(s)\geq 0$. Hence \eqref{vdot-rdot} holds in this case as well. 
        
        In either cases, it follows from \Cref{rem:vparaminM} that $v_p\leq \gamma_v(s)\leq v_q$ for every $s\in[0,1]$. Integrating \eqref{vdot-rdot} from $0$ to $s$ also gives 
        \begin{equation}
        \label{integrated-vdot-rdot}
        \gamma_r(s)\leq r_p+\frac{\gamma_v(s)-v_p}{2}\leq  r_p+\frac{v_q-v_p}{2}=:R.
    \end{equation}
    Since $[v_p,v_q]\Subset I$ and $m$ is continuous and positive, we set 
$$m_*\coloneqq \min_{v\in[v_p,v_q]}m(v)>0, \qquad \varepsilon\coloneqq \min\{r_q,m_*\}.$$
Whenever $\gamma_r(s)\leq \varepsilon$, we have
$$\gamma_r(s)\leq \varepsilon\leq m_*\leq m(\gamma_v(s))<2m(\gamma_v(s)).$$
Hence $\gamma(s)\in M_\mathrm{int}$ at every such parameter. More explicitly,
$$\mathcal F(\gamma_v(s),\gamma_r(s))\leq 1-\frac{2m_*}{\varepsilon}\leq -1,$$
so \eqref{vdot-rdot} implies 
\begin{equation}\label{ghinequality}
    \dot\gamma_r(s)\leq 0 \qquad \textrm{on any interval where}\qquad  \gamma_r(s)\leq \varepsilon.
\end{equation}

We claim that $\gamma_r(s)\geq\varepsilon$ for every $s\in[0,1]$. Suppose, for contradiction, that there exists $s_0\in[0,1)$ such that $\gamma_r(s_0)<\varepsilon$. Since $\gamma_r(1)=r_q\geq \varepsilon$, continuity guarantees a subsequent parameter where $\gamma_r(s)=\varepsilon$. Define $s_1\coloneqq \min\{s\in[s_0,1]:\gamma_r(s)=\varepsilon\}$. Then $s_1>s_0$ and $\gamma_r(s)<\varepsilon$ for every $s\in[s_0,s_1)$. Hence, we can use \eqref{ghinequality} to obtain the following.
$$\varepsilon=\gamma_r(s_1)=\gamma_r(s_0)+\int_{s_0}^{s_1}\dot\gamma_r(s)\ ds\leq \gamma_r(s_0)<\varepsilon,$$
which is a contradiction. Hence $\varepsilon\leq \gamma_r(s)\leq R$ for every $s\in[0,1]$. Together with \eqref{integrated-vdot-rdot}, this shows that 

\begin{figure}
    \centering
    \includegraphics[width=0.5\linewidth]{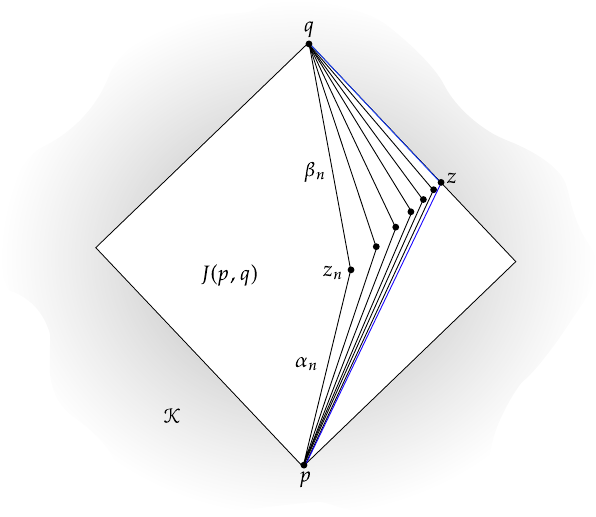}
    \caption{Compactness of the causal diamond $J(p,q)$ follows from a limit-curve argument.}
    \label{fig:compact-diamonds}
\end{figure}

\begin{equation}
    \label{in K}
    \gamma([0,1])\subseteq\mathcal{K}\coloneqq [v_p,v_q]\times [\varepsilon,R]\times \mathbb{S}^2\Subset M.
\end{equation} 
Since $\gamma$ was arbitrary, we have $J(p,q)\subseteq\mathcal{K}$. 

It remains to prove that $J(p,q)$ is closed in $\mathcal{K}$. Let $z_n\in J(p,q)$ with $z_n\to z\in \mathcal{K}$. If $z=p$ or $z=q$, then $z\in J(p,q)$ immediately. Otherwise, if $z_n\to z$ and $z\notin\{p,q\}$, there exists $N\in\mathbb{N}$ such that $z_n\notin\{p,q\}$ for every $n\geq N$. For each $n\geq N$, choose future-directed causal curves $\alpha_n,\beta_n:[0,1]\to J(p,q)$ with $\alpha_n(0)=p,\alpha_n(1)=z_n=\beta_n(0)$ and $\beta_n(1)=q$. Their concatenation connects $p$ to $q$, so \eqref{in K} gives $\alpha_n([0,1])\subseteq \mathcal{K}$ and $\beta_n([0,1])\subseteq\mathcal K$ for every $n\geq N$. We now apply Remark 2.54 of \cite{Minguzzi:2019mbe}: \emph{for causal curves contained in a compact set, non-total imprisonment excludes Item (ii) of Theorem 2.53 of \cite{Minguzzi:2019mbe}. Consequently, when their endpoints converge to distinct points, there exists a limit causal curve joining these points.} Applying this to $\alpha_n$, whose initial endpoints are $p$ and whose final endpoints converge to $z$, yields a future-directed causal curve from $p$ to $z$. On the other hand, applying it to $\beta_n$, whose initial endpoints converge to $z$ and whose final endpoints are $q$, yields a future-directed causal curve from $z$ to $q$. Hence $z\in J(p,q)$. It follows that $J(p,q)$ is closed in the compact set $\mathcal{K}$, and hence is compact. This completes the proof.

\end{proof}

The next proposition establishes a property of the Vaidya interior that obstructs the existence of any future $C^0$-extension through the curvature singularity $\{r=0\}$.
\begin{proposition}[$C^0$-obstruction]\label{obstruction}
Let $(M,g)$ be an ingoing Vaidya spacetime satisfying $m'(v)\geq 0$. For $\mu>0$, let $\gamma:[-\mu,0)\to M_{\mathrm{int}}$ be a future-directed future-inextendible timelike curve such that $r\circ\gamma(s)\to 0$ as $s\to 0$. There exists a sequence of spacelike Cauchy hypersurfaces $\{\Sigma_n\}_{n\in \mathbb{N}}$ of $(M_{\mathrm{int}},g)$ with $\Sigma_n\to\{r=0\}$ as $n\to \infty$ (locally uniformly as graphs, in the sense made clear in the proof below), along with a sequence of pairs of points $p_n,q_n\in I^{+}(\gamma(-\mu),M_{\mathrm{int}})\cap\Sigma_n$ such that
\begin{equation}
\label{eq:disst-blow}
   \lim_{n\to\infty} d_{\Sigma_n}(p_n,q_n)=\infty,
\end{equation}
i.e., the Riemannian distance between $p_n$ and $q_n$ as measured along $\Sigma_n$ diverges as one approaches the curvature singularity.
\end{proposition}
\begin{proof}
    \underline{\emph{Step 1: Construction of Cauchy hypersurfaces.}} Let $\psi:I\to \mathbb{R}$ be an increasing diffeomorphism and define a $C^2$ function $\tau:M_{\mathrm{int}}\to \mathbb{R}$ by
    \begin{equation}
        \label{Cauchy-temporal-fn}
        \tau(v,r,\omega)\coloneqq \psi(v) + H(v,r); \qquad H(v,r) \coloneqq -\mathrm{log}\left(\frac{r}{2m(v)-r}\right)
    \end{equation}
    We show that for $c\in \mathbb{R}$, the level sets $\Sigma_c\coloneqq \tau^{-1}(c)$ are Cauchy hypersurfaces of $M_{\mathrm{int}}$. To this end, let $\sigma:(s_-,s_+)\to M_{\mathrm{int}},\ \sigma(s) = (\sigma_v(s),\sigma_r(s),\sigma_\omega(s))$ be an inextendible future-directed timelike curve. First, by \Cref{prop:timefunction} we have $\dot\sigma_v(s)>0$ and $\dot\sigma_r(s)<0$. From the definition of $H$ in \eqref{Cauchy-temporal-fn} we have
  $$\partial_vH(v,r) = \frac{2m'(v)}{2m(v)-r}\geq 0, \qquad \partial_rH(v,r)=-\frac{1}{r}-\frac{1}{2m(v)-r}=-\frac{2m(v)}{r(2m(v)-r)}<0.$$
    Consequently we have 
    \begin{equation}
        \label{dH}
        \frac{d}{ds}H(\sigma(s)) = \frac{2m'(\sigma_v(s))}{2m(\sigma_v(s))-\sigma_r(s)} \dot\sigma_v(s)-\frac{2m(\sigma_v(s))}{\sigma_r(s)(2m(\sigma_v(s))-\sigma_r(s))}\dot\sigma_r(s)>0.
    \end{equation}
    It follows that $\tau\circ\sigma$ is strictly increasing. In what follows, we may fix some $s_0\in(s_-,s_+)$ and consider the restrictions $\sigma|_{[s_0,s_+)}$ and $\sigma|_{(s_-,s_0]}$ in order to study the behaviour of $\tau\circ\sigma$ at the future end $s\to s+$ and the past end $s\to s_-$ respectively.
    
    We claim that $\tau\circ\sigma(s)\to \infty$ as $s \to s_+$. Suppose, for contradiction, that $\tau\circ\sigma$ is bounded above on $[s_0,s_+)$. Since $H\circ\sigma$ is strictly increasing, we have $H(\sigma(s))> H(\sigma(s_0))$ for all $s\in [s_0,s_+)$, and hence boundedness of $\tau=\psi +H$ implies that $\psi\circ\sigma_v$ is bounded above. Since $\sigma_v$ is increasing and $\psi$ is an increasing diffeomorphism, $\sigma_v$ remains in a closed and bounded subinterval of $I$. Since $\sigma_r$ is strictly decreasing, there exists $r_+\geq 0$ such that $\sigma_r(s)\to r_+$ as $s\to s_+$. In the case $r_+=0$, we have $2m(\sigma_v(s))-\sigma_r(s)\geq 2m(\sigma_v(s_0))-\sigma_r(s_0)>0$ for every $s\geq s_0$, since $m'(v)\geq0$, $\sigma_v$ is increasing and $\sigma_r$ is decreasing. Therefore
    $$\lim_{s\to s_+}H\circ\sigma(s)=-\lim_{s\to s_+}\mathrm{log}\left(\frac{\sigma_r(s)}{2m(\sigma_v(s))-\sigma_r(s)}\right)=\infty,$$
    contradicting the boundedness of $\tau\circ\sigma$. Now, consider the case $r_+>0$. Reparameterising $\sigma|_{[s_0,s_+)}$ by $-r$, a calculation similar to the one in the proof of \Cref{lemma:vel-bounds} shows that $\sigma_v$ and $\sigma_\omega$ converge at the future end. Let $v_+\coloneqq \lim_{s\to s_+}\sigma_v(s)$ and $\omega_+\coloneqq \lim_{s\to s_+}\sigma_\omega(s)$. Together with the convergence of $\sigma_r$, this gives a future endpoint $(v_+,r_+,\omega_+)\in M_{\mathrm{int}}$ of $\sigma$, contradicting its future-inextendibility. Hence $\tau\circ\sigma(s)\to\infty$ as $s\to s_+$. 

    Similarly, we claim that $\tau\circ\sigma(s)\to -\infty $ as $s\to s_-$. Suppose instead that $\tau\circ\sigma$ is bounded below at the past endpoint. Since $H\circ\sigma$ is strictly increasing we have $H(\sigma(s))< H(\sigma(s_0))$ for all $s\leq s_0$. The lower bound for $\tau=\psi+H$ therefore gives a lower bound for $\psi\circ\sigma_v$. Since $\sigma_v(s)\leq \sigma_v(s_0)$ for all $s\leq s_0$, it follows that $\sigma_v$ remains in a closed and bounded subinterval of $I$. Towards the past, $\sigma_r$ is increasing. Since $0<\sigma_r(s)<2m(\sigma_v(s))$ and $\sigma_v$ remains in a closed and bounded interval, we have that $\sigma_r$ has a finite positive limit $r_->0$. Reparameterizing $\sigma|_{(s_-,s_0]}$ by $-r$, again a calculation similar to the one in the proof of \Cref{lemma:vel-bounds} shows that $\sigma_v$ and $\sigma_\omega$ converge at the past end. Set $v_-:=\lim_{s\to s_-}\sigma_v(s)$ and $\omega_-:=\lim_{s\to s_-}\sigma_\omega(s).$ By continuity of $m$, we have $2m(\sigma_v(s))-\sigma_r(s)\to \delta_-:=2m(v_-)-r_-\geq 0.$ If $\delta_-=0$, since $r_->0$, we have
$$\lim_{s\to s_-}H\circ\sigma(s)=-\lim_{s\to s_-}\log\left(\frac{\sigma_r(s)} {2m(\sigma_v(s))-\sigma_r(s)}\right)=-\infty.$$
Since $\psi\circ\sigma_v$ remains bounded, this implies
$\tau\circ\sigma(s)\to-\infty$, contradicting the assumed lower
bound. Hence $\delta_->0$. It follows that $(v_-,r_-,\omega_-)\in M_{\mathrm{int}}$ is a past endpoint of \(\sigma\), contradicting past-inextendibility. Hence $\tau\circ\sigma(s)\to -\infty$ as $s\to s_-$. 

Hence we have obtained that $\tau\circ\sigma$ is strictly increasing from $-\infty$ to $\infty$. Therefore every inextendible timelike curve meets every level set $\Sigma_c$ exactly once. Hence each $\Sigma_c$ is a Cauchy hypersurface of $M_\mathrm{int}$. We now write $\Sigma_c$ explicitly in terms of the coordinates $r$ and $v$. The equation $\tau=c$ is given by,
\begin{equation}
\label{r-eqn}
    \frac{r}{2m(v)-r}= e^{\psi(v)-c}\implies r=r_c(v)\coloneqq \frac{2m(v)}{1+e^{c-\psi(v)}}.
\end{equation}
Hence the Cauchy hypersurfaces are graphs of the form $\{r=r_c(v)\}$.

\underline{\emph{Step 2: Distance blow-up at the curvature singularity.}} Let $\gamma$ be the curve as in the statement of the proposition, and let $\gamma(-\mu/2) =(v_0,r_0,\omega_0)$. Since $\gamma(-\mu)\ll\gamma(-\mu/2)$ and chronological futures are open, there exist $a,b\in I$ with $a<v_0<b$ such that $(v,r_0,\omega_0)\in I^{+}(\gamma(-\mu),M_{\mathrm{int}})$ for every $v\in[a,b]$. For $v\in [a,b]$ and $0<r<r_0$, the radial curve from $(v,r_0,\omega_0)$ to $(v,r,\omega_0)$ has a tangent proportional to $-\partial_r$, and so is future-directed null. The push-up property gives
\begin{equation}
\label{containment}
    [a,b]\times (0,r_0)\times\{\omega_0\} \subseteq I^{+}(\gamma(-\mu),M_{\mathrm{int}}).
\end{equation}
Choose $c_0\in \mathbb{R}$ such that $r_c(a)<r_0$ and $r_c(b)<r_0$ whenever $c\geq c_0$. Set $c_n\coloneqq c_0+n$ and $\Sigma_n \coloneqq \Sigma_{c_n}$. From the construction of these Cauchy hypersurfaces in Step 1, we can make precise the sense in which these hypersurfaces approach the curvature singularity $\{r=0\}$: for every compact $K\Subset I$, we have
$$0\leq\lim_{n\to\infty}\sup_{v\in K}r_{c_n}(v)=\lim_{n\to\infty} \sup_{v\in K}\frac{2m(v)}{1+e^{c_n-\psi(v)}}\leq \lim_{n\to\infty} \left( 2\sup_{v\in K}m(v)e^{\sup_{v\in K}\psi(v)-c_n} \right)=0,$$
and hence
$$ \forall K\Subset I:\lim_{n\to \infty} \sup_{v\in K} |r_{c_n}(v)-0|=0,$$
i.e., $\Sigma_n\to \{r=0\}$ locally uniformly as graphs.

\begin{figure}
    \centering
    \includegraphics[width=0.6\linewidth]{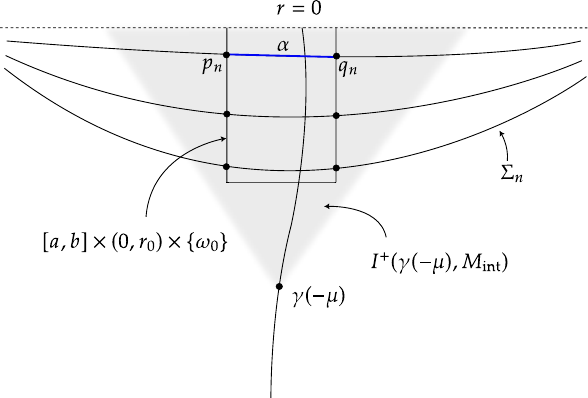}
    \caption{Distance-blow up at the curvature singularity for ingoing Vaidya spacetimes.}
    \label{fig:obstruction}
\end{figure}
Now define the sequence of points $p_n\coloneqq (a,r_{c_n}(a),\omega_0)$ and $q_n\coloneqq(b,r_{c_n}(b),\omega_0)$. By the inclusion \eqref{containment}, we have $p_n,q_n \in I^{+}(\gamma(-\mu),M_\mathrm{int})$. It remains to estimate their distance as measured along each $\Sigma_{c_n}$. Differentiating \eqref{r-eqn} gives
\begin{equation}
    \label{r-diff-eqn}
    r_c'(v) = \frac{2m'(v)}{1+e^{c-\psi(v)}}+\frac{2m(v)\psi'(v)e^{c-\psi(v)}}{(1+e^{c-\psi(v)})^2}>0.
\end{equation}
The induced metric on $\Sigma_c$ is 
\begin{equation}
    \label{induced-metric}
    g_c = \left(e^{c-\psi(v)}+2r_c'(v)\right) dv\otimes dv+r_c(v)^2 g_{\mathbb{S}^2}.
\end{equation}
By \eqref{r-diff-eqn}, the coefficient $e^{c-\psi(v)}+2r_c'(v)$ is strictly positive, and therefore $g_c$ is a Riemannian metric. In particular, $\Sigma_c$ is spacelike, which establishes that $\tau=\psi+H$ is actually a Cauchy temporal function on $M_\mathrm{int}$.

Let $\alpha:[0,1]\to \Sigma_n$ be any piecewise $C^1$ curve from $p_n$ to $q_n$ given by $\alpha(s) = (\alpha_v(s),r_{c_n}(\alpha_v(s)),\alpha_\omega(s))$. Then $\alpha_v(0)=a$ and $\alpha_v(1)=b$. Using \eqref{r-diff-eqn} and \eqref{induced-metric}, we obtain
$$L_{g_{c_n}}(\alpha)\geq e^{c_n/2}\int_0^1 e^{-\psi(\alpha_v(s))/2}|\dot\alpha_v(s)|\ ds\geq e^{c_n/2}\left|\int_0^1e^{-\psi(\alpha_v(s))/2 }\dot\alpha_v(s) \ ds\right|=e^{c_n/2}\int_a^b e^{-\psi(v)/2}\ dv.$$

Taking the infimum over all piecewise $C^1$ curves in $\Sigma_n$ from $p_n$ to $q_n$, we find that 
$$ d_{\Sigma_n}(p_n,q_n)\geq e^{c_n/2} \int_a^b e^{-\psi(v)/2}\ dv .$$
Taking the limit $n\to \infty$ for which  $c_n\to \infty$, we get the required distance blow-up. This completes the proof.

\end{proof}

Recall that a spacetime is called future one-connected if and only if any two future-directed timelike curves with the same endpoints are homotopic through future-directed timelike curves, with endpoints fixed throughout the homotopy.

\begin{proposition}[Future one-connectedness]
\label{prop:foc}
The interior $(M_{\mathrm{int}},g)$ of any ingoing Vaidya spacetime
satisfying $m'(v)\geq 0$ is future one-connected.
\end{proposition}

\begin{proof}
Let $p=(v_p,r_p,\omega_p)\ll q=(v_q,r_q,\omega_q)\in M_{\mathrm{int}}$ and for $i\in\{1,2\},$ let $\gamma_i:[r_q,r_p]\to M_{\mathrm{int}}$ with  $\gamma_i(r)=\bigl((\gamma_i)_v(r),r,(\gamma_i)_\omega(r)\bigr)$, be past-directed timelike curves satisfying $ \gamma_i(r_q)=q$ and $\gamma_i(r_p)=p$. This parametrisation is permitted by \Cref{prop:timefunction}. We need to show that
there exists a timelike homotopy with fixed endpoints between $\gamma_1$ and $\gamma_2$. The timelike inequality for $\gamma_i$, followed by completion of the square\footnote{Completing the square means for $A>0,$ we have $Ax^2+2x=A\left(x+\frac{1}{A}\right)^2-\frac{1}{A}$, and hence $Ax^2+2x\geq-\frac{1}{A}$.} in $(\dot\gamma_i)_v(r)$, gives
\begin{align*}
0>g\bigl(\dot\gamma_i(r),\dot\gamma_i(r)\bigr)&= \left( \frac{2m((\gamma_i)_v(r))}{r}-1\right)\bigl((\dot\gamma_i)_v(r)\bigr)^2+2(\dot\gamma_i)_v(r) +r^2 \bigl\|(\dot\gamma_i)_\omega(r)\bigr\|_{\mathbb S^2}^2 \\
&\geq-\frac{r}{2m((\gamma_i)_v(r))-r}+r^2\bigl\|(\dot\gamma_i)_\omega(r)\bigr\|_{\mathbb S^2}^2.
\end{align*}
It follows that
$$\bigl\|(\dot\gamma_i)_\omega(r)\bigr\|_{\mathbb S^2}<\frac{1}{\sqrt{r\bigl(2m((\gamma_i)_v(r))-r\bigr)}}.$$

Set $m_p\coloneqq m(v_p).$ Since $v_p\leq(\gamma_i)_v(r)\leq v_q$ and $m'(v)\geq0$, we have $m((\gamma_i)_v(r))\geq m_p.$ Moreover, $p\in M_{\mathrm{int}}$ gives $r_p<2m_p$. It follows that
\begin{equation}
\label{37}
    L\bigl((\gamma_i)_\omega\bigr)<\int_{r_q}^{r_p}\frac{\mathrm dr}{\sqrt{r(2m_p-r)}}<\int_0^{r_p}\frac{\mathrm dr}{\sqrt{r(2m_p-r)}} =2\sin^{-1}\sqrt{\frac{r_p}{2m_p}}<\pi.
\end{equation}

Here, the equality is obtained by making the substitution $r=2m_p\sin^2\alpha$ for $\alpha\in[0,\pi/2)$, and the final inequality is obtained since $r_p<2m_p$. Consequently, $\operatorname{Im}\bigl((\gamma_i)_\omega\bigr)\subseteq\mathbb S^2\setminus\{-\omega_q\},$ because every point of $(\gamma_i)_\omega$ has spherical distance less than $\pi$ from $\omega_q=(\gamma_i)_\omega(r_q)$. Consider the exponential map $\exp_{\omega_q}:B_\pi(0)\subset T_{\omega_q}\mathbb S^2\to\mathbb S^2$ based at $\omega_q$. For $i\in\{1,2\}$, define the \emph{angular straightening homotopy} $\Gamma_i:[r_q,r_p]\times[r_q,r_p]\to M_{\mathrm{int}}$ by

\begin{equation}
\label{38}
\Gamma_i(u,r) \coloneqq \begin{cases} \Bigl( (\gamma_i)_v(r),\,r,\, \exp_{\omega_q}\!\left( \ell_i(u,r) \exp_{\omega_q}^{-1}\bigl((\gamma_i)_\omega(u)\bigr) \right) \Bigr),
& r_q\leq r\leq u,\\[0.4em] \gamma_i(r), & u\leq r\leq r_p,\end{cases}
\end{equation}
where
$$\ell_i(u,r)\coloneqq\frac{\displaystyle\int_{r_q}^{r}\bigl\|(\dot\gamma_i)_\omega(r')\bigr\|_{\mathbb S^2}\,\mathrm dr'}{\displaystyle\int_{r_q}^{u}\bigl\|(\dot\gamma_i)_\omega(r')\bigr\|_{\mathbb S^2} \,\mathrm dr'}.
$$
If the denominator vanishes, then $(\gamma_i)_\omega$ is constant on $[r_q,u]$, and in this case, we set $\ell_i(u,r)=0$. With this convention, $\Gamma_i$ is continuous. At $r=u$, the two formulas in \eqref{38} agree. If the denominator of $\ell_i$ tends to zero, then
$$
\begin{aligned}
    \|\ell_i(u,r)\exp_{\omega_q}^{-1}((\gamma_i)_\omega(u))\|_{\mathbb{S}^2} &=\ell_i(u,r) \|\exp_{\omega_q}^{-1}((\gamma_i)_\omega(u))\|_{\mathbb{S}^2}\\ &\leq d_{\mathbb S^2}\bigl(\omega_q,(\gamma_i)_\omega(u)\bigr) \\&
     \leq \int_{r_q}^{u}\|(\dot\gamma_i)_\omega(r')\|_{\mathbb S^2}\,dr'\longrightarrow0.
\end{aligned}
$$

\begin{figure}
    \centering
    \includegraphics[width=0.35\linewidth]{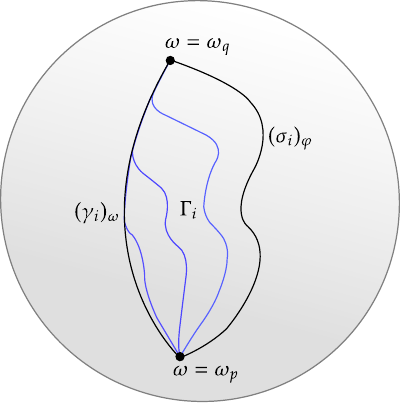}
    \caption{The angular straightening homotopy.}
    \label{fig:foc}
\end{figure}

We compute, for $r_q\leq r\leq u$ and at every regular point,
$$
\begin{aligned}
\small
g(\partial_r\Gamma_i(u,r),\partial_r\Gamma_i(u,r))&=-\mathcal F\bigl((\gamma_i)_v(r),r\bigr)(\dot\gamma_i)_v(r)^2+2(\dot\gamma_i)_v(r)+r^2\left\|\partial_r\!\left(\exp_{\omega_q}\!\left(\ell_i(u,r)\exp_{\omega_q}^{-1}\bigl((\gamma_i)_\omega(u)\bigr)\right)\right)\right\|_{\mathbb S^2}^{2}\\
&=-\mathcal F\bigl((\gamma_i)_v(r),r\bigr)(\dot\gamma_i)_v(r)^2+2(\dot\gamma_i)_v(r)+r^2|\partial_r\ell_i(u,r)|^2\left\|\exp_{\omega_q}^{-1}\bigl((\gamma_i)_\omega(u)\bigr)\right\|_{\mathbb S^2}^{2}\\
&= -\mathcal F((\gamma_i)_v(r),r)) \dot{(\gamma_i)}_v(r)^2 + 2 \dot{(\gamma_i)}_v(r)+r^2\|\dot{(\gamma_i)}_\omega(r)\|^2_{\mathbb{S}^2}\cdot \underbrace{\frac{d_{\mathbb S^2}^2((\gamma_i)_\omega(u),(\gamma_i)_\omega(r_q))}{\displaystyle \left(\int_{r_q}^u \|\dot{(\gamma_i)}_\omega(r')\|_{\mathbb S^2}\ dr'\right)^2}}_{\leq 1}\\
&\leq g\bigl(\dot\gamma_i(r),\dot\gamma_i(r)\bigr) <0.
\end{aligned}
$$

Here $\mathcal F(v,r)$ is the function defined in \eqref{vaidya-original-metric}. Hence $\Gamma_i$ is a timelike homotopy with fixed endpoints between $\gamma_i:[r_q,r_p]\to M_\mathrm{int}$ and the timelike curve $\sigma_i\coloneqq\Gamma_i(r_p,\cdot),$ whose projection on $\mathbb S^2$ lies on the unique minimizing geodesic arc connecting $\omega_p$ with $\omega_q$. We now introduce coordinates on $\mathbb{S}^2$ such that the geodesic $\mathbb S^1$ through $\omega_p$ and $\omega_q$ is parameterized by $\varphi\in(0,2\pi)$. It follows that $\sigma_i$ maps into the submanifold $N\coloneqq \{(v,r,\varphi)\in I\times (0,\infty)\mathbb \times \mathbb S^1:r<2m(v)\} \subseteq M_\mathrm{int}$, which carries the induced Lorentzian metric 
$$
g_N\coloneqq -\mathcal F(v,r) \ dv\otimes dv + (dv\otimes dr +dr \otimes dv)+r^2 d\varphi\otimes d\varphi.
$$ 
We now reverse the orientation of $\sigma_i$ and reparameterize it by $v$. Retaining the notation $\sigma_i$ for the resulting curve, we have $\sigma_i(v)=(v,(\sigma_i)_r(v),(\sigma_i)_\varphi(v))$. By timelikeness of $\sigma_i$ we have
$$0>g_N\bigl(\dot{\sigma}_i(v),\dot{\sigma}_i(v)\bigr)=-\left(1-\frac{2m(v)}{(\sigma_i)_r(v)}\right)+2(\dot{\sigma}_i)_r(v)+(\sigma_i)_r(v)^2
(\dot{\sigma}_i)_\varphi(v)^2.$$
Defining the \emph{reciprocal radius} $y\coloneqq 1/r$, dividing by $(\sigma_i)_r(v)^2>0$ gives
$$
\begin{aligned}
0&>-\frac{1}{(\sigma_i)_r(v)^2}+\frac{2m(v)}{(\sigma_i)_r(v)^3}+\frac{2(\dot{\sigma}_i)_r(v)}{(\sigma_i)_r(v)^2}+\bigl((\dot{\sigma}_i)_\varphi(v)\bigr)^2\\
&=-(\sigma_i)_y(v)^2+2m(v)(\sigma_i)_y(v)^3-2(\dot{\sigma}_i)_y(v)+\bigl((\dot{\sigma}_i)_\varphi(v)\bigr)^2.
\end{aligned}
$$
Hence we have
\begin{equation}
    \label{39}
    (\dot{\sigma}_i)_y(v)>m(v)(\sigma_i)_y(v)^3-\frac12(\sigma_i)_y(v)^2+\frac12\bigl((\dot{\sigma}_i)_\varphi(v)\bigr)^2.
\end{equation}
For each $v\in[v_p,v_q]$, and every positive function $h:[v_p,v_q]\to(0,\infty)$, we define a functional
$$G\circ h(v)\coloneqq m(v)h(v)^3-\frac12h(v)^2 .$$
The timelike inequality for $\sigma_i$ given by \eqref{39} may be rewritten in terms of this functional as
\begin{equation}
    \label{use}
    (\dot{\sigma}_i)_y(v)>G\circ(\sigma_i)_y(v)+\frac12\bigl((\dot{\sigma}_i)_\varphi(v)\bigr)^2.
\end{equation}

The following convexity argument is pointwise in $v$: Let $v\in[v_p,v_q]$ be arbitrary. Since $(\sigma_i)_r(v)<2m(v)$,
we have $(\sigma_i)_y(v)>1/(2m(v))$ for each $i\in\{1,2\}$. Hence every $y$ in the interval joining $(\sigma_1)_y(v)$ and $(\sigma_2)_y(v)$ satisfies $y>1/(2m(v))$, and therefore
$$\frac{d^2}{dy^2}\left(m(v)y^3-\frac12y^2\right)=6m(v)y-1>3-1=2>0.$$
Hence for each fixed $v\in[v_p,v_q]$, the map $y\mapsto m(v)y^3-(1/2)y^2$ is convex on the interval joining $(\sigma_1)_y(v)$ and $(\sigma_2)_y(v)$. 
Now, for $u\in[0,1]$, define the convex combinations
\begin{equation}\label{fosc:convexitydefn}
    y_u(v)\coloneqq(1-u)(\sigma_1)_y(v)+u(\sigma_2)_y(v),\qquad \varphi_u(v)\coloneqq(1-u)(\sigma_1)_\varphi(v)+u(\sigma_2)_\varphi(v).
\end{equation}
Then, in the $(v,y,\varphi)$ coordinates, define the homotopy 
$$
\Gamma:[0,1]\times [v_p,v_q]\to N~:~\Gamma(u,v)\coloneqq(v,y_u(v),\varphi_u(v)).
$$
We have
$$g_N=-\bigl(1-2m(v)y\bigr)\,dv\otimes dv-\frac{1}{y^2}(dv\otimes dy+dy\otimes dv)+\frac{1}{y^2}\,d\varphi\otimes d\varphi.$$
For the longitudinal curves $v\mapsto \Gamma(u,v)$, we compute pointwise
$$
\begin{aligned}
g_N\bigl(\partial_v\Gamma(u,v),\partial_v\Gamma(u,v)\bigr)&=-\bigl(1-2m(v)y_u(v)\bigr)-\frac{2\dot y_u(v)}{y_u(v)^2}+\frac{\dot\varphi_u(v)^2}{y_u(v)^2}\\
&=\frac{2m(v)y_u(v)^3-y_u(v)^2-2\dot y_u(v)+\dot\varphi_u(v)^2}{y_u(v)^2}\\
&=\frac{2}{y_u(v)^2}\left[G\circ y_u(v)-\dot y_u(v)+\frac12\dot\varphi_u(v)^2\right].
\end{aligned}
$$
Using the convexity $y\mapsto m(v)y^3-(1/2)y^2$ pointwise at each $v$, along with \eqref{fosc:convexitydefn} and the convexity $x\mapsto x^2$, we have
$$G\circ y_u(v)\leq(1-u)G\circ(\sigma_1)_y(v)+uG\circ(\sigma_2)_y(v), \qquad \dot\varphi_u(v)^2\leq(1-u)\bigl((\dot\sigma_1)_\varphi(v)\bigr)^2+u\bigl((\dot\sigma_2)_\varphi(v)\bigr)^2.$$
Substituting these pointwise estimates and using \eqref{use}, we compute
$$
\begin{aligned}
g_N\bigl(\partial_v\Gamma(u,v),\partial_v\Gamma(u,v)\bigr)&\leq\frac{2}{y_u(v)^2}\Bigg[(1-u)\cdot G\circ(\sigma_1)_y(v)+u\cdot G\circ(\sigma_2)_y(v)-(1-u)(\dot\sigma_1)_y(v)-u(\dot\sigma_2)_y(v)\\
&\qquad \qquad \qquad+\frac12\left((1-u)\bigl((\dot\sigma_1)_\varphi(v)\bigr)^2+u\bigl((\dot\sigma_2)_\varphi(v)\bigr)^2\right)\Bigg]
\\
&=\frac{2}{y_u(v)^2}\Bigg[(1-u)\left(G\circ(\sigma_1)_y(v)-(\dot\sigma_1)_y(v)+\frac12\bigl((\dot\sigma_1)_\varphi(v)\bigr)^2\right)\\
&\qquad\qquad \qquad+u\left(G\circ(\sigma_2)_y(v)-(\dot\sigma_2)_y(v)+\frac12\bigl((\dot\sigma_2)_\varphi(v)\bigr)^2\right)\Bigg] <0.
\end{aligned}
$$

Hence every longitudinal curve $v\mapsto\Gamma(u,v)$ is timelike. Moreover it is future-directed since $dv(\partial_v\Gamma(u,v))=1>0$. Hence $\Gamma$ is a fixed-endpoint timelike homotopy between $\sigma_1$ and $\sigma_2$ via future-directed timelike curves. This completes the proof.

\end{proof}

Finally, the next proposition establishes an important property of $M_\mathrm{int}$ that will be used in the proof of \Cref{thm:main thm3}. Before stating the proposition, we introduce \emph{mass-normalized coordinates}: for $v_*\in I$, we define
\begin{equation}
\label{mass-coord}
    t\coloneqq t(v)\coloneqq \int_{v_*}^v \frac{d\bar v}{m(\bar v)}, \qquad \rho\coloneqq \frac{r}{m(v)}.
\end{equation}
Since $t'(v)=1/m(v)>0$ and $r=m(v)\rho$, in these coordinates we have $M_\mathrm{int} = t(I)\times(0,2)\times \mathbb{S}^2$.
\begin{proposition}[Local preservation of timelikeness and time-orientation]
\label{preserving}
Let $D_0\subset t(I)$ be a compact interval, and let $C>0$. Then there exists $\epsilon_0>0$ and $\rho_0>0$ such that for every 
$$
1<\kappa<1+\epsilon_0, \qquad |c|<C(\kappa-1), \qquad \textrm{and}\qquad h\in \mathrm{SO}(3),
$$
the map 
\begin{equation}
    \label{map}
    \Phi_{\kappa,c,h}(t,\rho,\omega)\coloneqq (\kappa t+c,\kappa\rho,h(\omega))
\end{equation}
is defined on $D_0\times(0,\rho_0)\times \mathbb{S}^2$, takes values in $M_{\textrm{int}}$, and its differential maps every timelike vector to a timelike vector with the same time orientation. 
\end{proposition}
\begin{proof}
From \eqref{mass-coord}, we have $dt=dv/m(v)$ and $r=m(v)\rho$, and hence $dv=m(v) dt$ and $dr=m(v)(d\rho+\rho ~m'(v)dt).$ Writing $v=v(t)$ and $a(t)\coloneqq m'(v(t))$, substitution into the Vaidya metric \eqref{vaidya-original-metric} gives $g=m(v(t))^2 \hat g$, where
\begin{equation}
    \label{ghat}
    \hat g =\left(\frac{2}{\rho}-1+2a(t)\rho\right)dt\otimes dt+(dt\otimes d\rho +d\rho\otimes dt)+\rho^2 g_{\mathbb{S}^2}.
\end{equation}
The conformal factor is strictly positive, so it suffices to work with $(M_\mathrm{int},\hat g)$, equipped with the same time-orientation.

Now choose a compact interval $D\subset t(I)$ such that $D_0\subset \operatorname{Int}{(D)}$, and set 
$$T_0\coloneqq \sup_{t\in D_0} |t|,\qquad d_D\coloneqq \mathrm{dist}(D_0,\mathbb{R}\setminus \operatorname{Int}{(D)})=\inf\{|t-s|:t\in D_0,s\in \mathbb{R}\setminus\operatorname{Int}{(D)}\}>0.$$
Choose 
$$0<\epsilon_0<\min\left\{1,\frac{d_D}{T_0+C}\right\}.$$
Let $\epsilon\coloneqq \kappa-1\in(0,\epsilon_0)$ and $|c|<C\epsilon$. For every $t\in D_0$, we have 
$$
|(\kappa t+c)-t|=|\epsilon t+c|\leq |\epsilon t|+|c| =\epsilon|t|+|c|\leq(T_0+C)\epsilon\leq d_D,
$$
and hence $\kappa t+c\in \operatorname{Int}{(D)}\subset t(I)$. Additionally, $\sup_{s\in D}|a(s)|<\infty$, and $\sup_{s\in D}|a'(s)|<\infty$, since $a'(t)=m"(v(t))v'(t)=m"(v(t))m(v(t))$ is continuous, and $D$ is compact. 
Let $C_a\coloneqq 1+\sup_{t\in D} |a(t)|+2(T_0+C)\sup_{t\in D} |a'(t)|>0$. It follows that
$$
\begin{aligned}
|\kappa a(\kappa t+c)-a(t)|&=|(\kappa-1)a(t)+\kappa(a(\kappa t+c)-a(t))|\\
&=|\epsilon a(t)+\kappa a'(\xi)(\epsilon t+c)|\\
&\le \epsilon|a(t)|+\kappa|\epsilon t+c|\,|a'(\xi)|\\
&\le \epsilon\sup_{s\in D}|a(s)|+\kappa(\epsilon|t|+|c|)\sup_{s\in D}|a'(s)|\\
&\le \epsilon\sup_{s\in D}|a(s)|+\kappa(\epsilon T_0+C\epsilon)\sup_{s\in D}|a'(s)|\\
&\le \epsilon\sup_{s\in D}|a(s)|+2(T_0+C)\epsilon\sup_{s\in D}|a'(s)|\\
&=\epsilon(\sup_{s\in D}|a(s)|+2(T_0+C)\sup_{s\in D}|a'(s)|)\\
&=(C_a-1)\epsilon <C_a\epsilon.
\end{aligned}
$$
The second step above uses the mean value theorem, which gives $\xi\in(\min\{t,\kappa t+c\},\max\{t,\kappa t+c\})$.

Now, choose $0<\rho_0<1/2$ so small that $4C_a\rho_0^2<1$. If $0<\rho<\rho_0$, then $0<\kappa\rho<2\rho_0<1$. Together with $\kappa t+c\in D\subset t(I)$, this proves that the map $\Phi_{\kappa,c,h}$ defined in \eqref{map} maps $D_0\times(0,\rho_0)\times \mathbb{S}^2$ into $D\times(0,2\rho_0)\times \mathbb S^2 \subseteq M_\mathrm{int}$. Moreover, the same coordinate formula as in \eqref{map} defines a differentiable map
on an open neighbourhood of $D_0\times(0,\rho_0)\times\mathbb S^2$ in $M_{\mathrm{int}}$. Hence $d\Phi_{\kappa,c,h}|_p$ is defined on
$T_pM_{\mathrm{int}}$ for every $p\in D_0\times(0,\rho_0)\times\mathbb S^2$.
\begin{figure}
    \centering
    \includegraphics[width=0.5\linewidth]{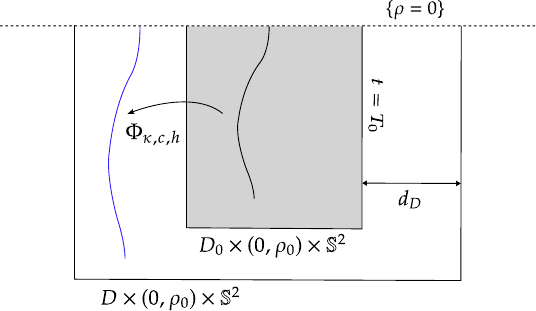}
    \caption{Preservation of timelikeness and time-orientation in a small region near the curvature singularity $\{\rho=0\}$.}
    \label{fig:prop2.14}
\end{figure}

Using $h^* g_{\mathbb S^2}=g_{\mathbb{S}^2}$, we compute
$$\begin{aligned}
    \Phi_{\kappa,c,h}^* \hat g&=\kappa^2\left(\frac{2}{\kappa\rho}-1+2\kappa\rho a(\kappa t+c)\right) dt\otimes dt+\kappa^2(dt\otimes d\rho+d\rho\otimes dt)+\kappa^2\rho^2 g_{\mathbb S^2}\\&=\kappa^2(\hat g+B_{\kappa,c}(t,\rho) dt\otimes dt),
\end{aligned}$$
where
$$B_{\kappa,c}(t,\rho)=-\frac{2(\kappa-1)}{\kappa\rho}+2\rho(\kappa a(\kappa t+c)-a(t)).$$
Consequently we have
$$B_{\kappa,c}(t,\rho)\leq -2\epsilon\frac{1}{\kappa\rho}+2C_a\epsilon\rho<-\frac{\epsilon}{\rho}+2C_a\epsilon\rho=\frac{\epsilon}{\rho}\left(-1+2C_a\rho^2\right)<-\frac{\epsilon}{2\rho}<0,$$
where the last estimate uses $2C_a\rho^2<2C_a\rho_0^2<1/2$. 
Now let $p\in D_0\times(0,\rho_0)\times \mathbb S^2$ and $X\in T_pM_\mathrm{int}$ be timelike. Then we have
$$\hat g_{\Phi_{\kappa,c,h}(p)}(d\Phi_{\kappa,c,h}|_p(X),d\Phi_{\kappa,c,h}|_p(X))=\kappa^2(\hat g_p(X,X)+B_{\kappa,c}(t_p,\rho_p)dt_p(X)^2)<0.$$
Hence the image vector is also timelike. As $dt=dv/m(v)$ and $m(v)>0$, future- and past-directed timelike vectors are characterized by $dt(X)>0$ and $dt(X)<0$ respectively.  We have
$$\ dt_{\Phi_{\kappa,c,h}(p)}(d\Phi_{\kappa,c,h}|_p(X))= \kappa dt_p(X),$$
Since $\kappa>0$, their signs are also the same, and hence the time-orientation of the image vector is also the same as $X$. The conclusion holds for $(M_\mathrm{int},g)$ since the Vaidya metric $g$ is conformal to $\hat g$. By the chain rule, these maps also take future-directed (resp. past-directed) timelike curves contained in $D_0\times (0,\rho_0)\times \mathbb S^2$ to future-directed (resp. past-directed) timelike curves. This completes the proof.

\end{proof}

\section{Proof of the main theorem}\label{sec-3}
We now proceed to prove the main \Cref{thm:main thm3} using arguments introduced for the Schwarzschild spacetime by Sbierski in \cite{Sbierski:2015nta, Sbierski_2018}. A conceptual overview of the proof and a comparison with the Schwarzschild case are given in \Cref{sec-4}. 

\begin{proof}[Proof of \Cref{thm:main thm3}]
        Assume that there exists a $C^0$-extension $(\tilde M,\tilde g)$ of $(M,g)$ with $\partial^+{\iota}(M) \neq \emptyset$. By \Cref{global-hyp}, $(M,g)$ is globally hyperbolic. Hence by Theorem 3.2 of \cite{Sbierski_2018}, we have that for any $\delta>0$ there exists a chart $\tilde\varphi:\tilde U\to \mathbb{R}_{\varepsilon_0,\varepsilon_1}\coloneqq (-\varepsilon_0,\varepsilon_0)\times (-\varepsilon_1,\varepsilon_1)^3$, along with a future-directed timelike geodesic $\tau:[-1,0) \to M$ which is future-inextendible in $M$, such that:
    \begin{enumerate}[(i)]
        \item $|\tilde g_{\mu\nu}(x) -\eta_{\mu\nu}|<\delta$ for all $x\in \mathbb R_{\varepsilon_0,\varepsilon_1}$.
        \item There exists a Lipschitz function $f:(-\varepsilon_1,\varepsilon_1)^3\to (-\varepsilon_0,\varepsilon_0)$ satisfying the inclusions 
        \begin{equation}\label{lipschitz-fn-graph}
            \operatorname{graph}f\coloneqq \{(x_0,\underline{x})\in \mathbb{R}_{\varepsilon_0,\varepsilon_1} : x_0 = f(\underline{x})\} \subseteq \tilde{\varphi}(\partial^{+}{\iota}(M)\cap\tilde{U})
        \end{equation}
        and
        \begin{equation}\label{lipschiztz-fn-below-graph}
            \operatorname{graph}f_{<} \coloneqq \{(x_0,\underline{x})\in \mathbb{R}_{\varepsilon_0,\varepsilon_1} : x_0 < f(\underline{x})\} \subseteq \tilde{\varphi}(\iota(M)\cap \tilde{U}).
        \end{equation}
Moreover, we have that $\operatorname{graph}f$ is achronal in $(\mathbb{R}_{\varepsilon_0,\varepsilon_1}, \tilde g)$.
        \item $\tilde\varphi\circ\iota\circ\tau \eqqcolon\bar \tau:[-1,0)\to \mathbb{R}_{\varepsilon_0,\varepsilon_1}$ maps into $\operatorname{graph}f_{<}$ and $\tilde\varphi\circ\iota\circ\tau(s)\to(0,0,0,0)$ as $s\to 0$. 
    \end{enumerate}
    Now fix some $a\in(0,1)$. Let $\langle\cdot,\cdot\rangle_{\mathbb{R}^4}$ denote the standard Euclidean inner-product on $\mathbb{R}^4$, and $|\cdot|_{\mathbb{R}^4}$ the associated standard Euclidean norm. We define
    \begin{equation}\label{cone-def}
        \begin{aligned}
              C^+_a &\coloneqq \left\{X\in\mathbb{R}^4:\frac{\langle X,e_0\rangle_{\mathbb{R}^4}}{|X|_{\mathbb{R}^4}} >a\right\},\\
        C^-_a &\coloneqq \left\{X\in\mathbb{R}^4:\frac{\langle X,e_0\rangle_{\mathbb{R}^4}}{|X|_{\mathbb{R}^4}} <-a\right\},\\
        C^c_a &\coloneqq \left\{X\in\mathbb{R}^4:-a<\frac{\langle X,e_0\rangle_{\mathbb{R}^4}}{|X|_{\mathbb{R}^4}} <a\right\}.\\
        \end{aligned}
    \end{equation}
    Here, $e_0$ is the unit (basis) vector along the $x_0$-axis. Clearly $C^+_a$ denotes the forward cone of vectors that make an angle less than $\mathrm{arccos}(a)$ with the $x_0$-axis, and $C^-_a$ is the corresponding backward cone. For Minkowski spacetime the chronological futures/pasts are given by $C^{\pm}_{1/\sqrt{2}}$. Since $5/8<1/\sqrt 2<5/6$, it follows that $C^+_{5/6}\subseteq C^+_{1/\sqrt 2}\subseteq C^+_{5/8}$, and hence we choose $0<\delta<1$ (which supplies the chart $(\tilde U,\tilde\varphi)$ above) so small that in $\mathbb{R}_{\varepsilon_0,\varepsilon_1}$, all the vectors in $C^+_{5/6}$ are future-directed timelike, all vectors in $C^-_{5/6}$ are past-directed timelike, and all vectors in $C^c_{5/8}$ are spacelike. It follows that for any $x\in\mathbb{R}_{\varepsilon_0,\varepsilon_1}$, one has
    \begin{equation}\label{cone-inclusions}
        \begin{aligned}
            (x+C^+_{5/6})\cap \mathbb{R}_{\varepsilon_0,\varepsilon_1} &\subseteq I^{+}(x,\mathbb{R}_{\varepsilon_0,\varepsilon_1})\subseteq (x+C^+_{5/8})\cap \mathbb{R}_{\varepsilon_0,\varepsilon_1},\\
        (x+C^-_{5/6})\cap \mathbb{R}_{\varepsilon_0,\varepsilon_1} &\subseteq I^{-}(x,\mathbb{R}_{\varepsilon_0,\varepsilon_1})\subseteq (x+C^-_{5/8})\cap \mathbb{R}_{\varepsilon_0,\varepsilon_1}.
        \end{aligned}
    \end{equation}
By  \Cref{timelike-incompletness}, one has that $\tau$ is either future complete (i.e., it leaves through the exterior) or $r\circ\tau(s)\to 0$ as $s\to 0$ (i.e., it leaves through the interior). Since $\dot{\bar\tau}(s)\in C^+_{5/8}$ for all $s\in[-1,0)$, one has $dx_0(\dot{\bar\tau}(s))>0$. Hence we may reparametrize $\bar \tau$ by $x_0$, i.e., we have $\bar\tau:[-s_0,0)\to \mathbb{R}_{\varepsilon_0,\varepsilon_1}$ given by $\bar\tau(s) = (s,\underline{\bar \tau}(s))$, where $-s_0\in (-\varepsilon_0,0)$. Since $\dot{\bar\tau}(s)\in C^+_{5/8}$ for all $s\in[-s_0,0)$, we have
$$\frac{5}{8}< \frac{\langle\dot{\bar\tau}(s),e_0\rangle_{\mathbb{R}^4}}{|\dot{\bar\tau}(s)|_{\mathbb{R}^4}} = \frac{1}{\sqrt{1+|\dot{\underline{\bar\tau}}(s)|^2}},$$ for all $s\in[-s_0,0)$. Combining this bound with the uniform upper bound on the metric components coming from $(i)$ gives for some $\xi>0$,
\begin{equation}
    \int_{-s_0}^0 \sqrt{-\tilde g(\dot{\bar \tau}(s),\dot{\bar\tau}(s))} \ ds = \int_{-s_0}^0 \sqrt{-\left(\tilde g_{00}+2\sum_{i=1}^3 \tilde g_{0i}\dot{\underline{\bar\tau}}^i(s)+\sum_{i,j=1}^3 \tilde g_{ij} \dot{\underline{\bar\tau}}^i(s)\dot{\underline{\bar\tau}}^j(s)\right)} \ ds \leq \xi <\infty.
\end{equation}
Hence $\tau$ has finite total proper time. Since it is a future-directed timelike geodesic in $M$, it is future-incomplete in $M$. \Cref{timelike-incompletness} then implies that $\tau$ cannot leave $M$ via the exterior, and that one has $r\circ\tau(s)\to 0$ as $s\to 0$. In what follows, we show that even this is not possible.

Consider the curve $\tilde\gamma:(-\varepsilon_0,0]\to\tilde M$ given in the chart $(\tilde U,\tilde\varphi)$ by $\tilde\varphi\circ\tilde\gamma(s) = (s,0,0,0)$. Define $\gamma\coloneqq\iota^{-1}\circ\tilde\gamma|_{(-\varepsilon_0,0)}$. Clearly $\gamma$ is a future-directed future-inextendible timelike curve in $M$. From Proposition 2.6 in \cite{Sbierski:2015nta} (see also Lemma 2.7 in \cite{Grant:2019rye}), $I^+(\tilde\gamma(s),\tilde U)$ is an open neighbourhood of $\tilde\gamma(0)$ for all $s\in(-\varepsilon_0,0)$. Hence there exists a future-directed timelike curve from $\tilde \gamma(s)$ to $\mathrm{Im}(\bar \tau)$ for all $s\in(-\varepsilon_0,0)$. By Item 1 of Remark 2.5 in \cite{Sbierski_2018}, this curve lies completely in $\iota(M)$. Consequently, for every $s<0$ and all sufficiently late parameters $t<0$, we have 
\begin{equation}
    \label{need}
    \gamma(s)\ll\tau(t)
\end{equation}
 in $M$. Since $v$ is strictly increasing along future-directed timelike curves in $M$ (by \Cref{rem:vparaminM}), it follows that $\gamma_v(s)<\tau_v(t)$. We claim that $r\circ\gamma(s)\to0$ as $s\to0$. Indeed, since $r\circ\tau(s)=\tau_r(s)\to0$ as $s\to0$, a final segment of $\tau$ lies in $M_\mathrm{int}$. Then \Cref{lemma:vel-bounds} shows that $v\circ\tau(s)=\tau_v(s)\to V<\infty$ as $s\to 0$. If $\sup I=\infty$, finiteness of $V$ already shows $V\in I$, and if $\sup I<\infty$, the argument in the finite-endpoint case of \Cref{timelike-incompletness} shows that $V\in I$. In either case we have $V\in I$. Moreover by \eqref{need}, we have $\gamma_v(s)\leq V$ for any $s$ in this final segment. Fix $a\in(-\varepsilon_0,0)$. Timelikeness gives $\dot\gamma_r(s)<\dot\gamma_v(s)/2$, so $\gamma_r -(\gamma_v/2)$ is decreasing and bounded below on $[a,0)$. Since $\gamma_v$ is increasing and bounded above, $\gamma_r(s)\to r_\infty\in[0,\infty)$ as $s\to 0$. If $r_\infty>0$, a final segment of $\gamma$ is contained in a compact subset  $[\gamma_v(a),V]\times[r_\infty/2,R]\times \mathbb S^2 \Subset M$ for some suitable $R$. This contradicts non-total imprisonment in the globally hyperbolic spacetime $(M,g)$. Hence $r_\infty=0$, which proves the claim. Positivity and monotonicity of the mass function $m$ then imply that a final segment of $\gamma$ lies in $M_\mathrm{int}$. For the remainder of the proof, we discard $\tau$ and work with $\gamma$.

 Now choose $0<\varepsilon_0'<\varepsilon_0$ such that $\gamma(-\varepsilon_0')\in M_{\mathrm{int}}$. By openness of $\iota(M_{\mathrm{int}})$ and continuity of $f$, with $f(0)=0$, we may choose $0<\varepsilon_1'<\varepsilon_1$ such that, for every $\underline{x}\in(-\varepsilon_1',\varepsilon_1')^3$,
$$ (-\varepsilon_0',\underline{x})\in\tilde\varphi(\iota(M_{\mathrm{int}})\cap\tilde U),\qquad |f(\underline{x})|<\varepsilon_0'.$$
The curve $x^0\mapsto\tilde\varphi^{-1}(x^0,\underline{x})$, $x^0\in[-\varepsilon_0',f(\underline{x})]$, is future-directed timelike. By \eqref{lipschiztz-fn-below-graph}, it lies in $\iota(M)$ before its endpoint, and hence in $\iota(M_{\mathrm{int}})$ by \emph{future invariance} of $M_\mathrm{int}$ \footnote{Recall that a subset $A$ of a spacetime $(M,g)$ is \emph{future invariant} under timelike curves if every future-directed timelike curve in $M$ starting in $A$ remains entirely in $A$, or equivalently, $I^+(A,M)\subseteq A$.}. By \eqref{lipschitz-fn-graph}, its endpoint lies outside $\iota(M)$ and therefore belongs to $\partial^+\iota(M_{\mathrm{int}})$. Thus, restricting $\tilde\varphi$ to $\tilde\varphi^{-1}(\mathbb R_{\varepsilon_0',\varepsilon_1'})$ and relabelling the parameters to $\varepsilon_0,\varepsilon_1$ shows that \eqref{lipschitz-fn-graph} and \eqref{lipschiztz-fn-below-graph} hold with $M$ replaced by $M_{\mathrm{int}}$.

 Note that after fixing $s'<0$, by \Cref{prop:timefunction}, we may reparameterise $\gamma|_{[s',0)}$ by $u=-r$ and write $\gamma(u)=(\gamma_v(u),-u,\gamma_\omega(u))$. This curve is future-directed timelike and has fixed past endpoint $\gamma(s')$. Hence \Cref{lemma:vel-bounds} applies. Returning to the original parameter $s$, we hence obtain $v_*\coloneqq \lim_{s\to 0} \gamma_v(s)$ and $\omega_* = \lim_{s\to0}\gamma_\omega(s)$. By \eqref{need} and the monotonicity of $v$ along future-directed timelike curves, we have $\gamma_v(s)\leq V$ for $s\in[s',0)$. Since $\gamma_v$ is increasing and $V\in I$, it follows that $v_*\in[\gamma_v(s'),V]\subset I$.

We switch now to the mass-normalised coordinates defined in \eqref{mass-coord} chosen with respect to the reference point $v_*$. We record two important estimates which will be used later twice. Let $p=(t_p,\rho_p,\omega_p)\ll q=(t_q,\rho_q,\omega_q)\in M_\mathrm{int}$ with $\rho_p<1$. Consider a past directed timelike curve $\beta$ from $q$ to $p$. Parametrizing it by $\rho$, we write $\beta(\rho)=(\beta_t(\rho),\rho,\beta_\omega(\rho))$ and note that it remains in the region $\{\rho<1\}$. Since $a(t)=m'(v(t))\geq 0$, the coefficient of $dt\otimes dt$ in the metric $\hat g$ given by \eqref{ghat} is at least $1/\rho$. With dots overhead now denoting differentiation in $\rho$, we have
$$\begin{aligned}
    0>\hat g(\dot\beta(\rho),\dot\beta(\rho))&\geq \frac{\dot\beta_t(\rho)^2}{\rho}+2\dot\beta_t(\rho)+\rho^2\|\dot\beta_\omega(\rho)\|^2_{\mathbb{S}^2}\\&=\frac{(\dot\beta_t(\rho)+\rho)^2}{\rho}+\rho^2\|\dot\beta_\omega(\rho)\|^2_{\mathbb{S}^2}-\rho,
\end{aligned}$$
for all $\rho\in[\rho_q,\rho_p]$. We now do a calculation similar to the one in the proof of \Cref{lemma:vel-bounds}: dropping the angular term in the first line gives $0<-\dot\beta_t(\rho)<2\rho$ and dropping the square in the second line yields $\|\dot\beta_\omega(\rho)\|_{\mathbb S^2}<\rho^{-1/2}$. Hence we have
\begin{equation}
    \label{tomega-conv}
    0<t_q-t_p=-\int_{\rho_q}^{\rho_p} \dot\beta_t(\rho) \ d\rho<\rho_p^2-\rho_q^2,\quad d_{\mathbb S^2}(\omega_p,\omega_q)\leq \int_{\rho_q}^{\rho_p} \|\dot\beta_\omega(\rho)\|_{\mathbb{S}^2}\ d\rho<2(\sqrt{\rho_p}-\sqrt{\rho_q}).
\end{equation}
These bounds are independent of the curve $\beta$ and require no additional mass bounds. Moreover, since $g=m(v(t))^2\hat g$ with a strictly positive conformal factor, $g$ and $\hat g$ have the same timelike curves with the chosen common time orientation. Hence the bounds in \eqref{tomega-conv} also apply to timelike curves of the original Vaidya metric $g$. Choose a compact interval $D_0\subset t(I)$ with $0\in\mathrm{Int}(D_0)$, and set 
\begin{equation}\label{mainthmT0def}
    T_0\coloneqq \sup_{t\in D_0} |t|\qquad \textrm{and}\qquad  \tilde C\coloneqq 1+T_0.
\end{equation}
Applying \Cref{preserving} with $C=\tilde C$, we denote the resulting constants by $\tilde \epsilon$ and $\tilde\rho$. We may choose them smaller, if necessary, so that both are less than $1$. Set $\mathcal S\coloneqq D_0\times (0,\tilde\rho)\times\mathbb S^2$ (this will be used later in steps 1.2.2 and 1.4). Finally choose $\max\{s',-\varepsilon_0\}<\tilde s<0$ so close to $0$ that $\gamma_t([\tilde s,0))\subseteq \operatorname{Int}(D_0)$ and $\gamma_\rho(\tilde s)<\tilde \rho$. Such a choice is possible since 
$$\gamma_t(s)=t(\gamma_v(s))\xrightarrow{s\to0} t(v_*)= 0,\qquad \gamma_\rho(s)=\frac{\gamma_r(s)}{m(\gamma_v(s))}\xrightarrow{s\to 0}\frac{0}{m(v_*)}= 0.$$ 
 Since $\gamma_\rho$ is strictly decreasing, it follows that $\gamma([\tilde s,0))\subseteq \mathcal S$. 

The rest of the proof proceeds in two steps. 

\smallskip
\textbf{Step $\mathbf 1$}. \textit{We show that there exists $\mu>0$ with,
\begin{align}
        \iota(I^+(\gamma(-\mu),M_{\mathrm{int}})) &\subseteq \tilde\varphi^{-1}(\mathrm{graph} f_<), \label{first}\\
        \left(-\varepsilon_0,-\frac{49}{50}\varepsilon_0\right)\times(-\varepsilon_1,\varepsilon_1)^3 &\subseteq I^{-}(\tilde\varphi\circ\tilde\gamma(-\mu), \mathbb{R}_{\varepsilon_0,\varepsilon_1}). \label{second}
    \end{align}}

\smallskip
\textbf{Step $\mathbf {1.1}$.} We can choose without loss of generality that $0<\varepsilon_1<\varepsilon_0/4$, by shrinking the chart. We now choose $x^+=(x_0^+,0,0,0)$ with $0<x_0^+<\varepsilon_0$ and $x^-=(x_0^-,0,0,0)$ with $\tilde s<x_0^-<0$ such that 
\begin{equation}
    \label{inc-1}
    \overline{(x^++C^-_{5/6})\cap(x^-+C^+_{5/6})}\Subset \mathbb{R}_{\varepsilon_0,\varepsilon_1}.
\end{equation}
These conditions can be imposed simultaneously: $x_0^+$ and $x_0^-$ may be chosen arbitrarily close to $0$, and the corresponding diamond then has compact closure in the chart. Moreover, we also have that the whole curve segment $\gamma([x_0^-,0))$ is contained in $\mathcal S$. Now choose $y^- \coloneqq (y_0^-,0,0,0)$ with $\max\{x_0^-,-\varepsilon_0/5\}<y_0^-<0$, so that
\begin{equation}
    \label{inc-2}
    \overline{C^-_{5/8}\cap(y^-+C^+_{5/8})}\Subset (x^++C^-_{6/7})\cap(x^-+C^+_{6/7}).
\end{equation}
Now let $z=(z^0,\underline{z}) \in (-\varepsilon_0,-49\varepsilon_0/50) \times (-\varepsilon_1,\varepsilon_1)^3$. Since $-\varepsilon_0/5<y_0^-$ and $\varepsilon_1<\varepsilon_0/4$, we have
$$(y^- -z)^0 = y_0^--z^0 > -\frac{\varepsilon_0}{5}+\frac{49}{50}\varepsilon_0 = \frac{39}{50}\varepsilon_0, \qquad |\underline{z}|_{\mathbb{R}^3}<\frac{\sqrt 3}{4}\varepsilon_0<\frac{\sqrt{11}}{5}(y^--z)^0.$$
By definition of $C^+_{5/6}$ (see \eqref{cone-def}), it follows that $y^--z\in C^+_{5/6}$, i.e., $z\in y^-+C^-_{5/6}$. From the inclusion \eqref{cone-inclusions}, we have $y^-+C^-_{5/6}\subseteq I^{-}(y^-,\mathbb{R}_{\varepsilon_0,\varepsilon_1})$, and so $z\in I^{-}(y^-,\mathbb{R}_{\varepsilon_0,\varepsilon_1})$. Since $z\in(-\varepsilon_0,-49\varepsilon_0/50)\times (-\varepsilon_1,\varepsilon_1)^3$ was chosen arbitrarily, we have that
\begin{equation}
    \label{second-1}
     \left(-\varepsilon_0,-\frac{49}{50}\varepsilon_0\right)\times(-\varepsilon_1,\varepsilon_1)^3 \subseteq I^{-}(y^-, \mathbb{R}_{\varepsilon_0,\varepsilon_1}).
\end{equation}

We next claim that, whenever $y_0^-<s<0$,
\begin{equation}
\label{equality of diamonds}
    \tilde\varphi^{-1}\left(I^{-}((s,0,0,0),\mathbb R_{\varepsilon_0,\varepsilon_1})\cap I^{+}(y^-,\mathbb{R}_{\varepsilon_0,\varepsilon_1})\right) = \iota \left(I^{-}(\gamma(s),M_{\mathrm{int}})\cap I^+(\gamma(y_0^-),M_\mathrm{int})\right).
\end{equation}
The inclusion ``$\subseteq$" follows from Item 1 of Remark 2.5 in \cite{Sbierski_2018}: if $\sigma$ is a past-directed timelike curve in $\mathbb{R}_{\varepsilon_0,\varepsilon_1}$ from $(s,0,0,0)$ to $y^-$, then $\mathrm{Im}(\tilde\varphi^{-1}\circ\sigma)\subseteq \iota(M_{\mathrm{int}})$ as $M_{\mathrm{int}}$ is preserved.

For the reverse inclusion ``$\supseteq$", let $\sigma:[y_0^-,s]\to M_\mathrm{int}$ be a future-directed timelike curve from $\gamma(y_0^-)$ to $\gamma(s)$. By \Cref{prop:foc}, there exists a timelike homotopy $\Gamma:[0,1] \times [y_0^-,s]\to M_\mathrm{int}$ with fixed endpoints between $\gamma|_{[y_0^-,s]}$ and $\sigma$. Hence $\iota\circ\Gamma$ is a timelike homotopy with fixed endpoints in $\tilde M$. It remains to show that $(\iota\circ\sigma)(\cdot) = (\iota\circ\Gamma)(1,\cdot)$ maps into $\tilde U$. We argue by continuity, showing that the set $J\coloneqq \{t\in[0,1]:\iota\circ\Gamma(t,[y_0^-,s])\subseteq \tilde U\}$ is non-empty, open and closed in $[0,1]$.
First, $0\in J$, since $\iota\circ\Gamma(0,\cdot)=\tilde\gamma|_{[y_0^-,s]}$. The openness of $J$ follows from the continuity of the homotopy, the compactness of $[y_0^-,s]$, and the openness of $\tilde U$. The closedness of $J$ follows from the fact that $I^-(\tilde\gamma(s),\tilde U)\cap I^+(\tilde\gamma(y_0^-),\tilde U)$ is precompact in $\tilde U$, so its closure in $\tilde M$ is contained in $\tilde U$. Connectedness of $[0,1]$ gives $J=[0,1]$, which in turn gives the desired conclusion, proving \eqref{equality of diamonds}.

Moreover, it follows from Proposition 2.7 of \cite{Sbierski:2015nta} that
$$I^{-}(0,\mathbb{R}_{\varepsilon_0,\varepsilon_1})\cap I^+(y^-,\mathbb R_{\varepsilon_0,\varepsilon_1})= \bigcup_{y_0^-<s<0} \left(I^-((s,0,0,0),\mathbb R_{\varepsilon_0,\varepsilon_1})\cap I^+(y^-,\mathbb R_{\varepsilon_0,\varepsilon_1})\right).$$
Combining this identity with \eqref{equality of diamonds}, we obtain
\begin{equation}
    \label{equality of diamonds 2}
    \tilde\varphi^{-1} \left(I^{-}(0,\mathbb{R}_{\varepsilon_0,\varepsilon_1})\cap I^+(y^-,\mathbb R_{\varepsilon_0,\varepsilon_1})\right) = \iota\left(\left(\bigcup_{y_0^-<s<0} I^-(\gamma(s), M_\mathrm{int})\right)\cap I^+(\gamma(y_0^-),M_\mathrm{int})\right).
\end{equation}
For the next step, recall that for a spacetime $(M,g)$, two sets $A,B\subseteq M$ are said to be \emph{timelike separated by a set $K\subseteq M$} iff every timelike curve from $A$ to $B$ intersects $K$. 

\smallskip
\textbf{Step $\mathbf{1.2.}$} For this step, we go back to the manifold $(M_{\mathrm{int}},g)$. Choose a $y_0^+$ with $y_0^-<y_0^+<0$ and define
\begin{equation}
\label{K-set}
    K\coloneqq \left[\left(\bigcup_{y_0^-<s<0} I^-(\gamma(s), M_\mathrm{int})\right)\cap I^+(\gamma(y_0^-),M_\mathrm{int})\right]\setminus I^+(\gamma(y_0^+), M_\mathrm{int})
\end{equation}

\smallskip
\textbf{Step $\mathbf{1.2.1.}$} We first show that $\gamma((y_0^+,0))$ and $I^-(\gamma(y_0^-),M_\mathrm{int})$ are timelike separated by $K$. To this end, let $\sigma:[0,1]\to M_\mathrm{int}$ be a past-directed timelike curve from $\sigma(0)\in \gamma((y_0^+,0))$ to $\sigma(1)\in I^-(\gamma(y_0^-), M_\mathrm{int})$. We first claim that there exists $\Delta_-\in(0,1)$ such that $\sigma^{-1}[I^+(\gamma(y_0^-),M_\mathrm{int})] = [0,\Delta_-)$.

Indeed, $0\in \sigma^{-1}[I^{+}(\gamma(y_0^-),M_\mathrm{int})]$. By the continuity of $\sigma$ and the openness of $I^+(\gamma(y_0^-),M_\mathrm{int})$, the preimage $\sigma^{-1}[I^{+}(\gamma(y_0^-),M_\mathrm{int})]$ is open in $[0,1]$. Furthermore, since $\sigma$ is past-directed timelike, whenever $s_0\in \sigma^{-1}[I^{+}(\gamma(y_0^-),M_\mathrm{int})]$, we also have $[0,s_0]\subseteq \sigma^{-1}[I^{+}(\gamma(y_0^-),M_\mathrm{int})]$. Finally, $(M_{\mathrm{int}},g)$ is chronological because it is globally hyperbolic (which follows from Step 1 of the proof of \Cref{obstruction}). Thus there are no closed timelike curves in $M_\mathrm{int}$, so $I^{-}(\gamma(y_0^-),M_\mathrm{int})$ is disjoint from $I^{+}(\gamma(y_0^-),M_\mathrm{int})$. Since the endpoint lies in the open chronological past, continuity places a final segment of the curve there as well. This gives $\Delta_-<1$.

\begin{figure}
    \centering
    \includegraphics[width=0.5\linewidth]{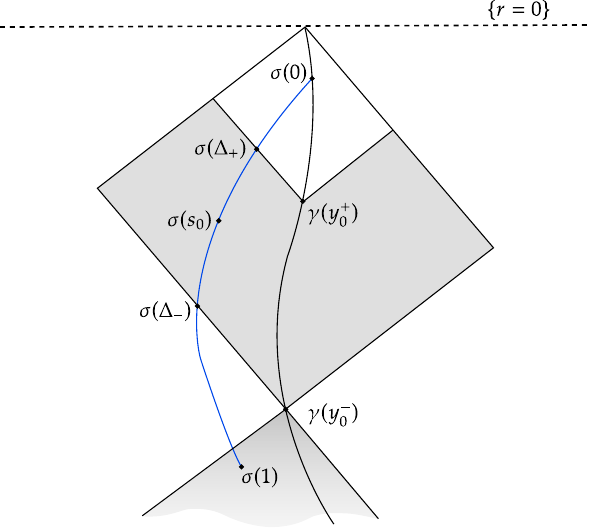}
    \caption{$K$ timelike separates $\gamma((y_0^+,0))$ and $I^-(\gamma(y_0^-),M_\mathrm{int})$.}
    \label{fig:step-1.2.1}
\end{figure}

A similar argument gives some $\Delta_+\in(0,1)$ with $\sigma^{-1}[I^{+}(\gamma(y_0^+),M_\mathrm{int})]=[0,\Delta_+)$. We next show that $\Delta_+<\Delta_-$. Since $(M_\mathrm{int},g)$ is globally hyperbolic and $g$ is $C^2$, we have $\overline{I^{+}(\gamma(y_0^+),M_\mathrm{int})} = J^{+}(\gamma(y_0^+),M_\mathrm{int})$. Combining this with $\gamma(y_0^+)\in I^{+}(\gamma(y_0^-),M_\mathrm{int})$, we obtain
$$\overline{I^{+}(\gamma(y_0^+),M_\mathrm{int})}=J^{+}(\gamma(y_0^+),M_\mathrm{int})\subseteq J^+\left(I^{+}(\gamma(y_0^-),M_\mathrm{int}),M_\mathrm{int}\right) = I^{+}(\gamma(y_0^-),M_\mathrm{int}).$$
It follows by the continuity of $\sigma$ that $\sigma(\Delta_+)\in I^{+}(\gamma(y_0^-),M_\mathrm{int})$, and hence $\Delta_+<\Delta_-$.

Now choose $s_0\in(\Delta_+,\Delta_-)$. Then $\sigma(s_0)\in I^{+}(\gamma(y_0^-),M_\mathrm{int})\setminus I^{+}(\gamma(y_0^+),M_\mathrm{int})$. Moreover, since the curve is past-directed timelike and starts on the prescribed segment of the reference curve, we have 
$$\sigma(s_0)\in \bigcup_{-\varepsilon_0<s<0} I^{-}(\gamma(s),M_\mathrm{int}),$$
and therefore $\sigma(s_0)\in K$. Hence $K$ timelike separates $\gamma((y_0^+,0))$ and $I^-(\gamma(y_0^-),M_\mathrm{int})$.

\smallskip
\textbf{Step $\mathbf{1.2.2.}$} Next we show that $K$ is precompact in $M_\mathrm{int}$. We first claim that for any $s_0\in(y_0^+,0)$, there exists $0<\epsilon<\tilde\epsilon$, a number $\delta>0$ and a neighbourhood $V$ of $\mathrm{id}\in \mathrm{SO}(3)$ such that, with $\kappa\coloneqq 1+\epsilon$, we have that
\begin{equation}
    \label{tube-in-future}
    \{\Phi_{\kappa,c,h}(\gamma(s)):|c|<\delta,h\in V\}\subseteq I^+(\gamma(y_0^+),M_\mathrm{int})
\end{equation}
holds for all $s\in(s_0,0)$.
Indeed, since $s_0>y_0^+$ and $\gamma$ is future-directed timelike, we have $ \gamma(s_0)\in I^+(\gamma(y_0^+),M_{\mathrm{int}}).$ Since $I^+(\gamma(y_0^+),M_{\mathrm{int}})$ is open and  $\Phi_{\kappa,0,\operatorname{id}}(\gamma(s_0))
 \to \gamma(s_0)$ as $\kappa\to1.$
There exists $\eta>0$ such that $\Phi_{\kappa,0,\operatorname{id}}(\gamma(s_0)) \in I^+(\gamma(y_0^+),M_{\mathrm{int}})$
whenever $1<\kappa<1+\eta$.  Choose $0<\epsilon<\min\{\eta,\tilde \epsilon\}$ and set $\kappa:=1+\epsilon$.
It follows that $\Phi_{\kappa,0,\operatorname{id}}(\gamma(s_0)) \in I^+(\gamma(y_0^+),M_{\mathrm{int}}).$
Now keep $\kappa$ fixed. The map  $(c,h)\mapsto\Phi_{\kappa,c,h}(\gamma(s_0))$ is continuous at $(0,\operatorname{id})$. Since $I^+(\gamma(y_0^+),M_{\mathrm{int}})$ is open, there exist $\delta'>0$ and a neighbourhood $V$ of $\operatorname{id}\in\mathrm{SO}(3)$ such that $\Phi_{\kappa,c,h}(\gamma(s_0))\in I^+(\gamma(y_0^+),M_{\mathrm{int}})$
for every $|c|<\delta'$ and every $h\in V$.  Define $\delta:=\min\{\delta',\tilde C\epsilon\}.$ Then the preceding inclusion continues to hold whenever $|c|<\delta$ and $h\in V$, and in addition
$|c|<\delta\leq \tilde C\epsilon=\tilde C(\kappa-1).$
Since $s_0>y_0^+>x_0^-$, we have $\gamma([s_0,0))\subseteq\mathcal S.$
Therefore, for every $|c|<\delta$ and $h\in V$, \Cref{preserving} shows that $\Phi_{\kappa,c,h}\circ\gamma|_{[s_0,0)}$ is future-directed timelike.  In particular, if $s\in(s_0,0)$, this curve gives a future-directed timelike curve from $\Phi_{\kappa,c,h}(\gamma(s_0))$ to $\Phi_{\kappa,c,h}(\gamma(s)).$ The first of these points already belongs to $I^+(\gamma(y_0^+),M_{\mathrm{int}})$. By transitivity of the
chronological relation, it follows that $\Phi_{\kappa,c,h}(\gamma(s))\in I^+(\gamma(y_0^+),M_{\mathrm{int}}).$ This proves \eqref{tube-in-future}.

\begin{figure}
    \centering
    \includegraphics[width=0.4\linewidth]{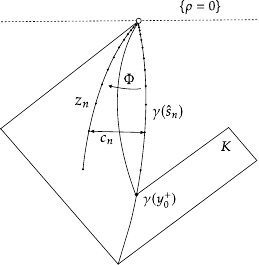}
    \caption{$K$ cannot approach $\{\rho=0\}$.}
    \label{fig:step-1.2.2}
\end{figure}

Now fix $s_0\in(y_0^+,0)$ and obtain $\epsilon,\delta$ and $V$ as in the claim. Suppose, for contradiction, that $K$ is not bounded away from $\rho=0$. Then there are points $z_n=(t_n,\rho_n,\omega_n)\in K$ with $\rho_n\to 0$. Choose $s_n\in(y_0^-,0)$ such that $z_n\in I^-(\gamma(s_n),M_\mathrm{int})$. Since $\rho$ strictly decreases towards the future, we have $0<\gamma_\rho(s_n)<\rho_n$ and hence $s_n\to 0$. Since $z_n\in I^+(\gamma(y_0^-), M_\mathrm{int})$, we also have $\rho_n<\gamma_\rho(y_0^-)<1$. Applying \Cref{tomega-conv} from $z_n$ to $\gamma(s_n)$ gives $0<\gamma_t(s_n)-t_n<\rho_n^2$ and $d_{\mathbb S^2}(\omega_n,\gamma_\omega(s_n))<2\sqrt{\rho_n}$. Consequently, $t_n\to 0$ and $\omega_n\to \omega_*$. For all sufficiently large $n$, let $\hat s_n\in(s_0,0)$ be determined by $\kappa\gamma_\rho(\hat s_n)=\rho_n$. Then $\hat s_n\to0$. Set $c_n\coloneqq t_n - \kappa\gamma_t(\hat s_n)$ and choose $h_n\in \mathrm{SO}(3)$ with $h_n(\gamma_\omega(\hat s_n))=\omega_n$. Since both angular points tend to $\omega_*$, the rotations may be chosen with angle tending to zero, so that $h_n\to\mathrm{id}\in \mathrm{SO}(3)$. Moreover, $c_n\to 0$. Hence for all sufficiently large $n$, $|c_n|<\delta$ and $h_n\in V$ we have $z_n=\Phi_{\kappa,c_n,h_n}(\gamma(\hat s_n))$. Equation \eqref{tube-in-future} then gives $z_n\in I^+(\gamma(y_0^+),M_\mathrm{int})$, contradicting the definition of $K$ in \eqref{K-set}. Hence there exists a $\rho_1>0$ such that $K\subseteq \{\rho\geq \rho_1\}$.

Finally the monotonicity of $t$ and $\rho$ gives $K\subseteq(\gamma_t(y_0^-),0)\times [\rho_1,\gamma_\rho(y_0^-))\times \mathbb S^2$. The closure of the set on the right is a compact subset of $M_\mathrm{int}$. This proves that $K$ is precompact in $M_\mathrm{int}$. 

\smallskip
\textbf{Step $\mathbf{1.3}$.} It follows from the definition of $K$ in \eqref{K-set}, together with \eqref{equality of diamonds 2}, that
\begin{equation}
    \label{incD}
    \overline{\tilde\varphi(\iota(K))}\subseteq \overline{I^-(0,\mathbb{R}_{\varepsilon_0,\varepsilon_1})\cap I^+(y_-,\mathbb{R}_{\varepsilon_0,\varepsilon_1})}.
\end{equation}
By \eqref{cone-inclusions}, \eqref{inc-1} and \eqref{inc-2}, we have
$$\overline{I^-(0,\mathbb{R}_{\varepsilon_0,\varepsilon_1})\cap I^+(y_-,\mathbb{R}_{\varepsilon_0,\varepsilon_1})}\subseteq \overline{C^-_{5/8}\cap(y^-+C^+_{5/8})}\Subset(x^++C^-_{6/7})\cap (x^-+C^+_{6/7})\Subset \mathbb{R}_{\varepsilon_0,\varepsilon_1}.$$ By continuity of $\tilde\varphi^{-1}$, we have that $\tilde\varphi^{-1}\bigl(\overline{I^-(0,\mathbb{R}_{\varepsilon_0,\varepsilon_1})\cap I^+(y_-,\mathbb{R}_{\varepsilon_0,\varepsilon_1})}\bigr)$ is compact, hence closed, in $\tilde M$, and so by \eqref{incD} we have
$$\overline{\iota(K)}\subseteq \tilde\varphi^{-1}\bigl(\overline{I^-(0,\mathbb{R}_{\varepsilon_0,\varepsilon_1})\cap I^+(y_-,\mathbb{R}_{\varepsilon_0,\varepsilon_1})}\bigr). $$
 By continuity of $\iota$, we have 
 $$
     \iota(\overline K) \subseteq \overline{\iota(K)} \subseteq \tilde\varphi^{-1}\bigl(\overline{I^-(0,\mathbb{R}_{\varepsilon_0,\varepsilon_1})\cap I^+(y_-,\mathbb{R}_{\varepsilon_0,\varepsilon_1})}\bigr) \subseteq \tilde\varphi^{-1}\left((x^++C^-_{6/7})\cap (x^-+C^+_{6/7})\right)
$$
  It follows that
$$ \mathcal W\coloneqq (\tilde\varphi\circ\iota)^{-1}\left((x^++C^-_{6/7})\cap (x^-+C^+_{6/7})\right)\subseteq M_\mathrm{int},$$
is an open neighbourhood of $\overline{K}$ in $M_\mathrm{int}$. 

\smallskip
\textbf{Step $\mathbf{1.4}$.} We show that there exists a $\mu>0$ such that $I^+(\gamma(-\mu),M_\mathrm{int})$ is timelike separated from $\{\gamma(x_0^-)\}$ by $\mathcal W$.

We consider $M_\mathrm{int}=t(I)\times(0,2)\times\mathbb S^2$ with the metric $d_\mathrm{M_{\mathrm{int}}}:M_\mathrm{int}\times M_\mathrm{int}\to [0,\infty)$ given by 
$$d_\mathrm{M_\mathrm{int}}((t_1,\rho_1,\omega_1),(t_2,\rho_2,\omega_2))\coloneqq |t_1-t_2|+|\rho_1-\rho_2|+d_{\mathbb S^2}(\omega_1,\omega_2).$$
 Since $\overline{K}$ is compact and contained in the open set $\mathcal W$, there exists a $\delta>0$ such that $K^\delta\coloneqq \{z\in M_\mathrm{int}:d_{M_\mathrm{int}}(z,\overline{K})<\delta\}\subseteq \mathcal W.$ Moreover by choosing $\delta$ smaller if necessary, we can also assume that 
 \begin{equation}
     \label{ball-contained}
     B_\delta(\gamma(x_0^-))\subseteq I^-(\gamma(y_0^-), M_\mathrm{int}).
 \end{equation} 

 \textbf{Step $\mathbf{1.4.1}$} We define a metric $d_{\mathrm{SO}(3)}:\mathrm{SO}(3)\times\mathrm{SO}(3)\to[0,\infty)$ by
 $$d_{\mathrm{SO}(3)}(f,h)\coloneqq \sup_{\omega\in\mathbb{S}^2}d_{\mathbb{S}^2}(f(\omega),h(\omega)).$$
 It is easy to see that 
 \begin{equation}
     \label{SO3-prop}
     \forall \eta>0:\forall\omega_0,\omega_1\in \mathbb{S}^2:d_{\mathbb{S}^2}(\omega_0,\omega_1)<\eta \implies\exists h\in\mathrm{SO}(3): d_{\mathrm{SO}(3)}(h,\mathrm{id})<\eta\ \text{and}\ h(\omega_0)=\omega_1.
 \end{equation}

Set $L\coloneqq \mathrm{diam}(D_0)+\gamma_\rho(x_0^-)$, where $\mathrm{diam}(D_0)\coloneqq \sup\{|t-s|:t,s\in D_0\}$. Choose $0<\epsilon<\tilde\epsilon$ (this is different from the $\epsilon$ in Step 1.2.2) so small that 
\begin{equation}\label{epsilon L lt deltaby4}
    \epsilon L<\frac \delta 4,
\end{equation}
and set $\kappa\coloneqq 1+\epsilon$. 

We continue to prove Step 1.4. For $0<\nu<-y_0^-$ and $p=(t_p,\rho_p,\omega_p)\in I^+(\gamma(-\nu),M_\mathrm{int})$, equation \eqref{tomega-conv} applied from $\gamma(-\nu)$ to $p$ gives
$$0<t_p-\gamma_t(-\nu)<\gamma_\rho(-\nu)^2,\qquad d_{\mathbb{S}^2}(\omega_p,\gamma_\omega(-\nu))<2\sqrt{\gamma_\rho(-\nu)}.$$
Both right-hand sides tend to zero as $\nu\to 0$, independently of $p$. Since $\gamma_t(-\nu)\to 0$ and $\gamma_\omega(-\nu)\to \omega_*$, the triangle inequalities for $|\cdot|$ and $d_{\mathbb{S}^2}$ yield
$$\begin{aligned}
   0 &\leq\sup_{p\in I^+(\gamma(-\nu),M_\mathrm{int})}\left(|t_p|+d_{\mathbb S^2}(\omega_p,\omega_*)\right)\\
   &\leq |\gamma_t(-\nu)|+d_{\mathbb{S}^2}(\gamma_\omega(-\nu),\omega_*)+\sup_{p\in I^+(\gamma(-\nu),M_\mathrm{int})}\left(|t_p-\gamma_t(-\nu)|+d_{\mathbb S^2}(\omega_p,\gamma_\omega(-\nu))\right)\\
   &\leq |\gamma_t(-\nu)|+d_{\mathbb S^2}(\gamma_\omega(-\nu),\omega_*)+\gamma_\rho(-\nu)^2+2\sqrt{\gamma_\rho(-\nu)} \xrightarrow{\nu\to0}0.
\end{aligned}$$
Thus, every point in the entire set $I^+(\gamma(-\nu), M_\mathrm{int})$  has $t_p$ uniformly close to $0$ and $\omega_p$ uniformly close to $\omega_*$ when $\nu$ is small.

Moreover, $0<\rho_p<\gamma_\rho(-\nu)$. Once $\kappa\gamma_\rho(-\nu)<\gamma_\rho(y_0^+)$, the strict monotonicity of $\gamma_\rho$ gives a unique $s_p\in(y_0^+,0)$ with $0<\gamma_\rho(s_p)=\kappa \rho_p<\kappa\gamma_\rho(-\nu)\to 0$ as $\nu\to 0$. This implies that $s_p\to 0$ uniformly over all such $p$. Hence $\gamma_t(s_p)\to 0$ and $\gamma_\omega(s_p)\to\omega_*$ uniformly as well. 

\begin{figure}
    \centering
    \includegraphics[width=0.5\linewidth]{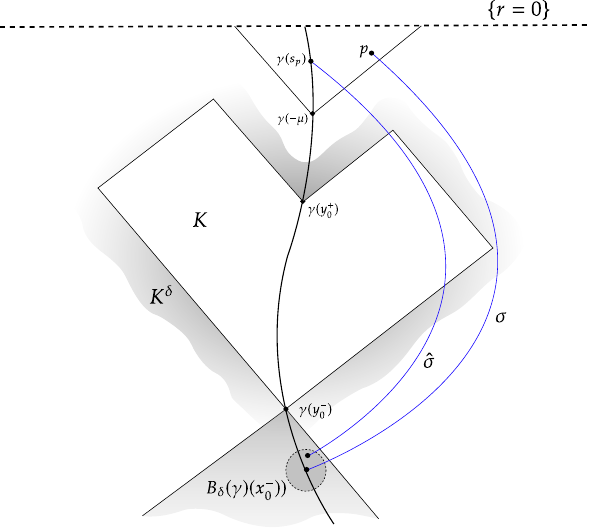}
    \caption{$K^\delta\subseteq \mathcal W$ timelike separates $I^+(\gamma(-\mu),M_\mathrm{int})$ and $\gamma(x_0^-)$.}
    \label{fig:step-1.4}
\end{figure}

After $\delta$ and $\epsilon$ have been fixed, we may therefore choose $\mu\in(0,-y_0^+)$ sufficiently small that $\kappa\gamma_\rho(-\mu)<\gamma_\rho(y_0^+)$ and, for every $p\in I^+(\gamma(-\mu),M_\mathrm{int})$, the corresponding point $s_p\in (y_0^+,0)$ satisfies
\begin{equation}
    \label{tomega-upperbds}
    |t_p-\gamma_t(s_p)|<\min\left\{\epsilon,\frac{\delta}{4}\right\},\qquad d_{\mathbb{S}^2}(\omega_p,\gamma_\omega(s_p))<\frac{\delta}{4},
\end{equation}
and $t_p,\gamma_t(s_p)\in \operatorname{Int}(D_0)$. 

In the following, we show that $I^+(\gamma(-\mu),M_\mathrm{int})$ is timelike separated from $\{\gamma(x_0^-)\}$ even by $K^\delta\subseteq W$, which, of course, would prove Step 1.4. So let $\sigma:[0,1]\to M_\mathrm{int}$ be a past-directed timelike curve with $\sigma(0)=p\in I^+(\gamma(-\mu),M_\mathrm{int})$ and $\sigma(1)=\gamma(x_0^-)$. By \eqref{SO3-prop} and \eqref{tomega-upperbds}, choose $h_p\in \mathrm{SO}(3)$ such that $d_{\mathrm{SO}(3)}(h_p,\mathrm{id})<\delta/4$ and $h_p(\omega_p)=\gamma_\omega(s_p)$. Set $c_p\coloneqq \gamma_t(s_p)-\kappa t_p$ and $\hat\sigma\coloneqq \Phi_{\kappa,c_p,h_p}\circ\sigma$. Then $\hat \sigma(0)=\gamma(s_p)$. Moreover by \Cref{mainthmT0def} and \eqref{tomega-upperbds}, we have 
$$
|c_p|\leq |\gamma_t(s_p)- t_p|+\epsilon|t_p|<\tilde C\epsilon.
$$
Along $\sigma$, the coordinate $t$ decreases and $\rho$ increases. Therefore
$\gamma_t(x_0^-)\leq \sigma_t(u)\leq t_p$ and $\rho_p\leq \sigma_\rho(u)\leq \gamma_\rho(x_0^-)$ for all $u\in[0,1]$, and so the whole curve $\sigma$ lies in $\mathcal S$. Together with $0<\epsilon<\tilde\epsilon$ and $|c_p|<\tilde C\epsilon$, \Cref{preserving} therefore shows that $\hat\sigma$ is past-directed timelike. 

Finally, for every $u\in[0,1]$, we have
\begin{equation}
    \label{dist-between-curves-bdd}
    \begin{aligned}
d_{M_{\mathrm{int}}}\bigl(\hat{\sigma}(u),\sigma(u)\bigr) &= |\kappa\sigma_t(u)+c_p-\sigma_t(u)|+|\kappa\sigma_{\rho}(u)-\sigma_{\rho}(u)|+d_{\mathbb{S}^2}(h_p(\sigma_\omega(u)),\sigma_\omega(u))\\
&=|\epsilon\sigma_t(u)+c_p|+\epsilon\sigma_\rho(u)+d_{\mathbb{S}^2}(h_p(\sigma_\omega(u)),\sigma_\omega(u)) \\
&= |\epsilon(\sigma_t(u)-t_p)+\gamma_t(s_p)-t_p|+\epsilon\sigma_\rho(u)+d_{\mathbb{S}^2}(h_p(\sigma_\omega(u)),\sigma_\omega(u)) \\
&\leq \epsilon(|\sigma_t(u)-t_p|+\sigma_\rho(u))+|\gamma_t(s_p)-t_p|+d_{\mathrm{SO}(3)}(h_p,\mathrm{id}) \\
&< \frac{\delta}{4}+\frac{\delta}{4}+\frac{\delta}{4}<\delta.
    \end{aligned}
\end{equation}
 Here, $|\sigma_t(u)-t_p|\leq \operatorname{diam}(D_0)$ and $\sigma_\rho(u)\leq \gamma_\rho({x}_0^-)$, so the first term is bounded by $\epsilon L<\delta/4$ (see \eqref{epsilon L lt deltaby4}), while each of the last two terms is smaller than $\delta/4$ by \eqref{tomega-upperbds} and the choice of $h_p$. In particular, \eqref{ball-contained} gives $\hat\sigma(1)\in I^-(\gamma(y_0^-),M_\mathrm{int})$. Since $\hat \sigma(0)=\gamma(s_p)$ with $s_p\in(y_0^+,0)$, Step 1.2.1 yields a $\hat u\in[0,1]$ with $\hat\sigma(\hat u)\in K$. The preceding distance estimate in \eqref{dist-between-curves-bdd} then gives $\sigma(\hat u)\in K^\delta\subseteq \mathcal W$, which concludes Step 1.4.

\smallskip
\textbf{Step 1.5.} First we show the following: Let $z:[0,1]\to \mathbb R_{\varepsilon_0,\varepsilon_1}$ be a continuous curve. If $z(0)\in \mathrm{graph} f_<$ and $z([0,1])\subset \tilde\varphi(\iota(M_{\mathrm{int}})\cap \tilde U)$, then $z([0,1])\subset \mathrm{graph} f_<$. Indeed, we may write the curve in coordinates as $z(s)=(z_0(s),\underline{z}(s))$, and define the continuous function $Z(s)\coloneqq z_0(s)-f(\underline{z}(s))$. Then $Z(0)<0$ since $z(0)<f(\underline{z}(0))$ by assumption. Now fix $s_1\in[0,1]$, and suppose $Z(s_1)=0$, i.e., $z(s_1)\in\mathrm{graph} f\subseteq \tilde\varphi(\partial^+ \iota(M_\mathrm{int})\cap \tilde U)$. This is a contradiction to the assumption that $z([0,1])\subset \tilde\varphi(\iota(M_\mathrm{int})\cap\tilde U)$. Hence $Z(s)\neq 0$ for any $s\in[0,1]$. Finally, if $Z(s)\geq 0$ for some $s_1\in[0,1]$, then the intermediate value theorem would give $Z(s_2)=0$ for  some $s_2\in[0,s_1]$, contradicting that $Z(s)\neq 0$ for any $s\in[0,1]$. Hence we have $Z(s)<0$ for all $s\in[0,1]$, or equivalently, $z([0,1])\subset \mathrm{graph} f_<$.

We now use the above result to show that
\begin{equation}
    \label{diamon-inc}
    \iota(I^+(\gamma(-\mu),M_{\mathrm{int}}))\subseteq\tilde\varphi^{-1}\bigl((x^++C^-_{6/7})\cap(x^-+C^+_{6/7})\bigr).
\end{equation}
The proof is by contradiction. Let $\sigma:[0,1]\to M_{\mathrm{int}}$ be a future-directed timelike curve with $\sigma(0)=\gamma(-\mu)$, and assume that for some $s'\in[0,1]$, we have $\iota\circ\sigma(s')\notin \tilde\varphi^{-1}
 \bigl((x^++C^-_{6/7})\cap(x^-+C^+_{6/7})\bigr)$.
Let
$$
 s_0:=\sup\bigl\{s\in[0,1]:
 \iota\circ\sigma(u)\in
 \tilde\varphi^{-1}
 \bigl((x^++C^-_{6/7})\cap(x^-+C^+_{6/7})\bigr)\
 \text{for all }u\in[0,s)\bigr\}.
$$
Clearly, $0<s_0\leq1$. Since $\overline{(x^++C^-_{6/7})\cap(x^-+C^+_{6/7})}\Subset \mathbb R_{\varepsilon_0,\varepsilon_1}$, continuity of $\iota\circ\sigma$ and the definition of $s_0$
give $(\tilde\varphi\circ\iota\circ\sigma)(s_0)\in
 \partial((x^++C^-_{6/7})\cap(x^-+C^+_{6/7})).$ 
It follows from the previous argument that $\tilde\varphi\circ\iota\circ\sigma(s_0)\in \mathrm{graph} f_<$. Now, since all vectors in $C^-_{5/6}$ are past-directed timelike, we can find a past-directed timelike curve $\tau:[0,1]\to \mathbb R_{\varepsilon_0,\varepsilon_1}$ with $\tau(0)=(\tilde\varphi\circ\iota\circ\sigma)(s_0)$ and $\tau(1)=x^-$,
which does not intersect $(x^++C^-_{6/7})\cap(x^-+C^+_{6/7}).$
Item~1 of Remark~2.5 in \cite{Sbierski_2018} implies that $\tau$ maps into
$\tilde\varphi(\iota(M_{\mathrm{int}})\cap\tilde U).$
Hence $(\tilde\varphi\circ\iota)^{-1}\circ\tau$ is a past-directed timelike curve in $M_{\mathrm{int}}$ with initial point $ \sigma(s_0)\in I^+(\gamma(-\mu),M_{\mathrm{int}})$ and final point $\gamma(x_0^-)$, which does not intersect $\mathcal W=(\tilde\varphi\circ\iota)^{-1}((x^++C^-_{6/7})\cap(x^-+C^+_{6/7})).$ This, however, is a contradiction to Step~1.4. Hence we have shown that
$$
 \iota(I^+(\gamma(-\mu),M_{\mathrm{int}}))\subseteq
 \tilde\varphi^{-1}\bigl((x^++C^-_{6/7})\cap(x^-+C^+_{6/7})\bigr),
$$
This concludes Step 1.5.

\begin{figure}
    \centering
    \includegraphics[width=0.5\linewidth]{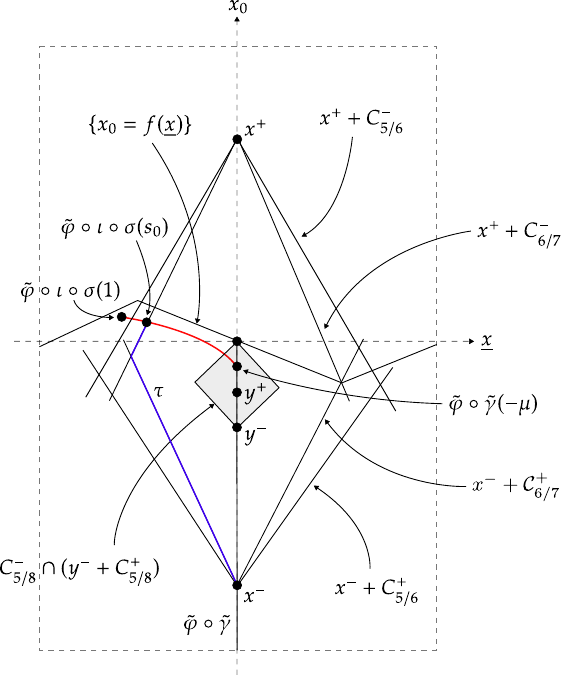}
    \caption{Proof of \eqref{first} in Step 1.}
    \label{fig:placeholder}
\end{figure}

\medskip 

We now finish the proof of Step 1. Let $p\in I^+(\gamma(-\mu),M_\mathrm{int})$ and let $\bar \sigma:[0,1]\to M_\mathrm{int}$ be a future-directed timelike curve with $\bar\sigma(0)=\gamma(-\mu)$ and $\bar\sigma(1)=p$. For every $s\in(0,1]$, we have $\bar\sigma(s)\in I^+(\gamma(-\mu),M_\mathrm{int})$. Hence \eqref{diamon-inc} together with $\iota\circ\gamma(-\mu)\in \tilde\varphi^{-1}\bigl((x^++C^-_{6/7})\cap(x^-+C^+_{6/7})\bigr)$ gives $$\iota\circ\bar\sigma([0,1])\subseteq\tilde\varphi^{-1}\bigl((x^++C^-_{6/7})\cap(x^-+C^+_{6/7})\bigr) \subset \tilde U.$$
It follows that $\tilde\varphi\circ\iota\circ\bar\sigma([0,1])\subset\tilde\varphi(\iota(M_\mathrm{int})\cap\tilde U)$ . Moreover $-\mu<0=f((0,0,0))$, hence $\tilde\varphi\circ\iota\circ\bar\sigma(0)=(-\mu,0,0,0)\in\mathrm{graph} f_<$. Again by the argument from Step 1.5, we have $\tilde\varphi\circ\iota\circ\bar\sigma(1)=\tilde\varphi(\iota(p))\in \mathrm{graph} f_<$, which proves \eqref{first} since $p\in I^+(\gamma(-\mu),M_\mathrm{int})$ was arbitrary. 

Moreover \eqref{second} also holds by \eqref{second-1}, along with the transitivity of the chronological relation since $y_0^-<y_0^+<-\mu<0$. This concludes the proof of Step~1.
\color{black}
\medskip

\textbf{Step $\mathbf{2}$.} \emph{We show that Step 1 guarantees the existence of a constant $C_d$ such that for any $C^1$ Cauchy hypersurface $\Sigma$ of $M_{\mathrm{int}}$ and any pair of points $p,q\in I^{+}(\gamma(-\mu),M_{\mathrm{int}})\cap\Sigma$, their distance in $\Sigma$ is bounded by $C_d$, i.e., $d_{\Sigma}(p,q)\leq C_d$.} Here $d_\Sigma$ denotes the intrinsic length pseudodistance induced by
$g|_{T\Sigma}$, taking the infimum over all piecewise-$C^1$ curves in
$\Sigma$ joining the given points. Since $\Sigma$ is achronal, its
induced tensor is nonnegative definite, but it may be degenerate.

Let $\Sigma$ be a $C^1$ Cauchy hypersurface. Assume $I^+(\gamma(-\mu),M_{\mathrm{int}})\cap\Sigma \neq \emptyset$, since otherwise there is nothing to prove. It follows that $\gamma(-\mu)\in I^-(\Sigma,M_{\mathrm{int}})$.

For every $\underline{x}\in(-\varepsilon_1,\varepsilon_1)^3$, consider the future-directed timelike curve
$$
\sigma_{\underline{x}}:(-\varepsilon_0,f(\underline{x}))\to M_{\mathrm{int}},
\qquad
\sigma_{\underline{x}}(s)=(\tilde\varphi\circ\iota)^{-1}(s,\underline{x}).
$$
Here $f$ is the Lipschitz function whose graph lies in $\tilde\varphi(\partial^+\iota(M_{\mathrm{int}})\cap\tilde U)$. As $s\to f(\underline{x})$, the image $\iota\circ\sigma_{\underline{x}}(s)$ converges to the boundary point $\tilde\varphi^{-1}(f(\underline{x}),\underline{x})$. Consequently, $\sigma_{\underline{x}}$ has no future endpoint in $M_{\mathrm{int}}$ and is future-inextendible there. By the inclusion \eqref{second}, each $\sigma_{\underline{x}}$ starts in $I^-(\gamma(-\mu),M_{\mathrm{int}})\subseteq I^-(\Sigma,M_{\mathrm{int}})$. Since $\Sigma$ is a Cauchy hypersurface, each $\sigma_{\underline{x}}$ meets $\Sigma$ exactly once. 

This defines a function $h:(-\varepsilon_1,\varepsilon_1)^3\to(-\varepsilon_0,\varepsilon_0)$ satisfying
$$
-\varepsilon_0<h(\underline{x})<f(\underline{x}),
\qquad
(h(\underline{x}),\underline{x})\in
\tilde\varphi(\iota(\Sigma)\cap\tilde U).
$$
Its graph is precisely the portion of $\tilde\varphi(\iota(\Sigma)\cap\tilde U)$ lying below the graph of $f$.

\smallskip

\textbf{Step $\mathbf{2.1}$.} Fix some $\underline{x}_0\in(-\varepsilon_1,\varepsilon_1)^3$. Since $\tilde\varphi(\iota(\Sigma)\cap\tilde U)$ is a $C^1$ embedded hypersurface, there exists an open neighbourhood $W\subseteq\mathbb{R}_{\varepsilon_0,\varepsilon_1}$ of $(h(\underline{x}_0),\underline{x}_0)$ and a $C^1$ submersion $\mathsf u:W\to\mathbb{R}$ such that $\tilde\varphi(\iota(\Sigma)\cap\tilde U)\cap W =\{\mathsf u=0\}$. Since $h(\underline{x}_0)<f(\underline{x}_0)$ and $f$ is continuous, we may shrink $W$ so that it lies entirely below the graph of $f$. Achronality of $\Sigma$ implies that the tangent space of $\Sigma$ contains no timelike vector. Since $\partial_0$ is timelike in the chart, it is not tangent to $\tilde\varphi(\iota(\Sigma)\cap\tilde U)$. The tangent space of the level set $\{\mathsf u=0\}$ is $\ker d\mathsf u$, and so $\partial_0\mathsf u|_{(h(\underline{x}_0),\underline{x}_0)}\neq0$. The implicit function theorem therefore gives an open neighbourhood $V\subseteq(-\varepsilon_1,\varepsilon_1)^3$ of $\underline{x}_0$ and a $C^1$ map $\mathsf v:V\to\mathbb{R}$ such that $(\mathsf v(\underline{x}),\underline{x})\in W$ and $\mathsf u((\mathsf v(\underline{x}),\underline{x}))=0$ for every $\underline{x}\in V$. Since $W$ lies below the graph of $f$, uniqueness of the intersection of each $\sigma_{\underline{x}}$ with $\Sigma$ gives $h|_V=\mathsf v$. As $\underline{x}_0$ was arbitrary, $h$ is $C^1$.

\smallskip

\textbf{Step $\mathbf{2.2}$.} For $i=1,2,3$, the vector $(\partial_i h)\partial_0+\partial_i$ is tangent to $\tilde\varphi(\iota(\Sigma)\cap\tilde U)$ at $(h(\underline{x}),\underline{x})$. Since this tangent space contains only non-timelike vectors, we have
\begin{equation}\label{slope-ineq}
    0\leq
    \tilde g((\partial_i h)\partial_0+\partial_i,
             (\partial_i h)\partial_0+\partial_i)
    =(\partial_i h)^2\tilde g_{00}
      +2(\partial_i h)\tilde g_{0i}
      +\tilde g_{ii}.
\end{equation}
Here the metric components are evaluated at $(h(\underline{x}),\underline{x})$. Recall that in Step 1 we chose $0<\delta<1$ and that $|\tilde g_{\mu\nu}(x)-\eta_{\mu\nu}|<\delta$ for all $x\in\mathbb{R}_{\varepsilon_0,\varepsilon_1}$. Hence we have

\begin{equation}
    \label{comp-bds}
    \tilde g_{00}<-(1-\delta)<0,\qquad
|\tilde g_{0i}|<\delta,\qquad
1-\delta<\tilde g_{ii}<1+\delta.
\end{equation}
It follows that equality in \eqref{slope-ineq} holds if and only if,
$$
(\partial_i h)_\pm
=\frac{-\tilde g_{0i}\mp
\sqrt{(\tilde g_{0i})^2-\tilde g_{00}\tilde g_{ii}}}
{\tilde g_{00}},
\qquad
(\partial_i h)_-<(\partial_i h)_+.
$$
The bounds \eqref{comp-bds} on the metric components imply
\begin{equation}
    \label{cslope}
    \max\{|(\partial_i h)_+|,|(\partial_i h)_-|\}
\leq
\frac{\delta+\sqrt{\delta^2+(1+\delta)^2}}{1-\delta}
\coloneqq C_{\mathrm{slope}}.
\end{equation}
The inequality \eqref{slope-ineq} along with $\tilde g_{00}<0$ gives $(\partial_i h)_-\leq\partial_i h\leq(\partial_i h)_+$, and hence $|\partial_i h|\leq C_{\mathrm{slope}}$ for $i=1,2,3$. Note that the constant $C_{\mathrm{slope}}$ in \eqref{cslope} depends only on $\delta$, and not on $h$ or $\Sigma$.

Define $\omega:(-\varepsilon_1,\varepsilon_1)^3\to\mathbb{R}_{\varepsilon_0,\varepsilon_1}$ by $\omega(\underline{x})\coloneqq(h(\underline{x}),\underline{x})$, i.e., $\omega$ parameterizes $C^1$ embedded hypersurface $\tilde S \coloneqq \tilde\varphi(\iota(\Sigma)\cap \tilde U)\cap \mathrm{graph} f_<\subseteq\mathbb{R}_{\varepsilon_0,\varepsilon_1}$. By continuity of $f$, we have that $\mathrm{graph} f_<$ is open in $\mathbb{R}_{\varepsilon_0,\varepsilon_1}$. It follows that $\tilde S$ is relatively open in $\tilde\varphi(\iota(\Sigma)\cap \tilde U)$, and so its preimage $S\coloneqq (\tilde\varphi\circ\iota)^{-1}(\tilde S)$ is open in $\Sigma$. Moreover, $(\tilde\varphi\circ\iota)^{-1}:\tilde S\to S$ is an isometry. The pullback $\bar g\coloneqq\omega^*\tilde g$ is a continuous, non-negative definite symmetric $2$-tensor on $(-\varepsilon_1,\varepsilon_1)^3$. It may be degenerate if $\Sigma$ has null tangent directions. For $i,j=1,2,3$, let $\bar g_{ij}$ denote its components with respect to the coordinates $\underline{x}$.

\begin{figure}[t]
\includegraphics[scale=0.8]{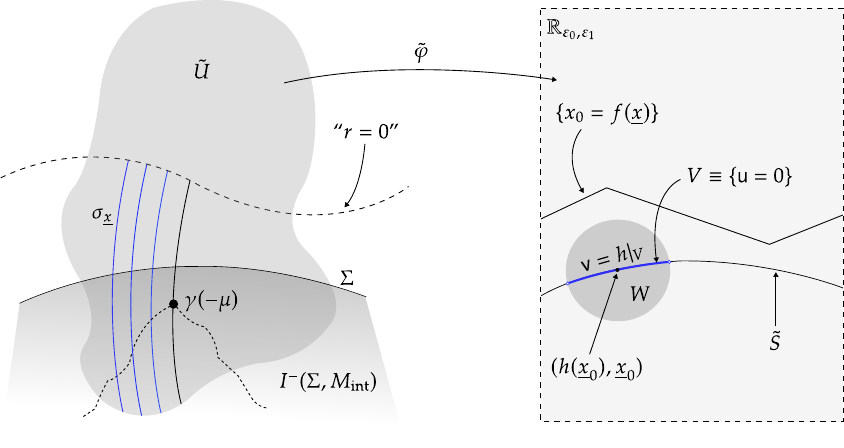}
\caption{The set-up of step 2.}
\end{figure}

\smallskip

\textbf{Step $\mathbf{2.3}$.} We compute
$$
\bar g_{ij}
=(\omega^*\tilde g)_{ij}
=\tilde g_{\mu\nu}
\frac{\partial\omega^\mu}{\partial x_i}
\frac{\partial\omega^\nu}{\partial x_j}
=\tilde g_{00}\frac{\partial h}{\partial x_i}
                    \frac{\partial h}{\partial x_j}
+\tilde g_{i0}\frac{\partial h}{\partial x_j}
+\tilde g_{0j}\frac{\partial h}{\partial x_i}
+\tilde g_{ij}.
$$
Using the metric bounds in \eqref{comp-bds} and the slope bounds from Step 2.2, we obtain
$$
|\bar g_{ij}|
\leq
(1+\delta)C_{\mathrm{slope}}^2
+2\delta C_{\mathrm{slope}}
+1+\delta
\coloneqq C_{\bar g},
\qquad i,j=1,2,3.
$$
Observe that $C_{\bar g}$ also depends only on $\delta$ and is independent of $\Sigma$.

Now fix any two points $p,q\in I^+(\gamma(-\mu),M_{\mathrm{int}})\cap\Sigma$. By \eqref{first} and the graph description above, there exist $\underline{x},\underline{y}\in(-\varepsilon_1,\varepsilon_1)^3$ such that $\omega(\underline{x})=\tilde\varphi\circ\iota(p)$ and $\omega(\underline{y})=\tilde\varphi\circ\iota(q)$. Define $\sigma:[0,1]\to(-\varepsilon_1,\varepsilon_1)^3$ by $\sigma(s)=\underline{x}+s(\underline{y}-\underline{x}).$ Its length with respect to $\bar g$ satisfies
\begin{align*}
    L(\sigma)=\int_0^1
      \sqrt{\bar g_{\sigma(s)}(\dot\sigma(s),\dot\sigma(s))}\,ds
    &=\int_0^1
      \sqrt{\sum_{i,j=1}^3
      (\underline{y}_i-\underline{x}_i)
      \bar g_{ij}(\sigma(s))
      (\underline{y}_j-\underline{x}_j)}\,ds\\
    &\leq\int_0^1
      \sqrt{\sum_{i,j=1}^3
      |\underline{y}_i-\underline{x}_i|\,
      C_{\bar g}\,
      |\underline{y}_j-\underline{x}_j|}\,ds\\
    &\leq\int_0^1
      \sqrt{3\cdot2\varepsilon_1\cdot C_{\bar g}
      \cdot3\cdot2\varepsilon_1}\,ds =6\varepsilon_1\sqrt{C_{\bar g}}.
\end{align*}
Here we used $|\underline{y}_i-\underline{x}_i|\leq|\underline{y}_i|+|\underline{x}_i|\leq2\varepsilon_1$ for $i=1,2,3.$ 

The piecewise-$C^1$ curve $(\tilde\varphi\circ\iota)^{-1}\circ\omega\circ\sigma: [0,1]\to S\subseteq\Sigma$ joins $p$ to $q$. The length of this curve in $\Sigma$ equals $L(\sigma)$, since $\bar g=\omega^*\tilde g$ and $\iota$ is an isometry. We therefore have
$$
d_\Sigma(p,q)\leq L(\sigma) \leq6\varepsilon_1\sqrt{C_{\bar g}}.
$$
Thus Step 2 holds with $C_d\coloneqq6\varepsilon_1\sqrt{C_{\bar g}}$, which is independent of $\Sigma$, $p$ and $q$.

\medskip

We now complete the proof of \Cref{thm:main thm3}. Applying \Cref{obstruction} with the fixed point $\gamma(-\mu)$ (coming from Step 1) gives a sequence of spacelike Cauchy hypersurfaces $\Sigma_n$ and points $p_n,q_n\in I^+(\gamma(-\mu),M_{\mathrm{int}})\cap\Sigma_n$ such that $d_{\Sigma_n}(p_n,q_n)\to\infty$.  Each $\Sigma_n$ is a $C^1$ Cauchy hypersurface, and its intrinsic length distance agrees with its Riemannian distance because it is spacelike. Step 2 therefore applies to every $\Sigma_n$, with the same constant $C_d$ as above, and yields $d_{\Sigma_n}(p_n,q_n)\leq C_d$ for every $n$. This is a contradiction. Hence the curve $\gamma$, and hence the geodesic $\tau$ cannot exit through the interior as well, from which it follows that $(M,g)$ is future $C^0$-inextendible.

\end{proof}

\section{Comparison with the Schwarzschild case}\label{sec-4}

We summarise the additional arguments that we need to extend Sbierski’s methods to Vaidya spacetimes, accounting for the varying mass and possible finite endpoints of $I$.

In \Cref{global-hyp}, we establish global hyperbolicity of ingoing Vaidya spacetimes as defined in \Cref{Vaidya-spacetime}, using the temporal function $T=v-r$ to establish strong causality, and subsequently proving the compactness of causal diamonds. This allows using Theorem 3.2 of \cite{Sbierski_2018}, which requires global hyperbolicity and establishes the existence of a Minkowskian chart centred at a boundary point.

In \Cref{timelike-incompletness}, we establish the timelike geodesic dichotomy for ingoing Vaidya spacetimes (satisfying the weak energy condition and the non-integrability condition in \Cref{non-integrability-condition}): a future-directed, future-inextendible timelike geodesic is incomplete if and only if $r\to0$. Its proof combines the existing result of \Cref{ThmA2} (due to Sbierski in \cite{Sbierski:2020toa}) with results from the Vaidya geodesic equations.

The $ C^0$-obstruction in \Cref{obstruction} requires a new construction of Cauchy hypersurfaces, carried out in Step~1 of its proof. In Schwarzschild spacetime, for example, constant-$r$ hypersurfaces are Cauchy hypersurfaces of the interior. In \Cref{obstruction}, we introduce a $C^2$ temporal function $\tau$, as defined in \eqref{Cauchy-temporal-fn}. Proving that $\tau$ strictly increases from $-\infty$ to $+\infty$ along every inextendible future-directed timelike curve ensures that its level sets are Cauchy hypersurfaces. Step 2 then establishes that there exists a sequence of pairs of points on the sequence of these Cauchy hypersurfaces such that distance diverges on these hypersurfaces.

To establish future one-connectedness of the interior of Schwarzschild spacetime close to the central singularity, two arbitrary timelike curves with fixed endpoints are parametrised by their corresponding radial coordinate $r$. The time-translation isometry $(t,r,\omega)\mapsto(t+c,r,\omega)$ of the Schwarzschild interior (in Schwarzschild coordinates) is then exploited to construct the timelike homotopy between these timelike curves. In Vaidya spacetimes, the dependence of $m$ on $v$ prevents this argument from applying directly. Despite this, in \Cref{prop:foc}, we establish future one-connectedness of its interior, provided the weak energy condition is satisfied.

In \Cref{preserving}, we provide the main replacement for the time-translation isometry $(t,r,\omega)\mapsto(t+c,r,\omega)$ used in Schwarzschild spacetime. In Vaidya spacetimes, due to the absence of this isometry, time translations generally do not preserve timelikeness.  In the mass-normalised coordinates~\eqref{mass-coord}, the map $\Phi_{\kappa,c,h}$ defined in \eqref{map}, preserves every timelike vector and its time orientation on a suitable region. This replacement is used to establish precompactness of the separating set $K$ defined in \eqref{K-set} in Step~1.2.2 of the proof of \Cref{thm:main thm3}. Additionally, it is also used to  prove the statement in Step~1.4 of \Cref{thm:main thm3}.

\section*{Declaration}
During the preparation of the manuscript, the authors benefited from occasional discussions with an AI for parts of \Cref{obstruction}, \Cref{prop:foc}, and \Cref{preserving}. AI was also used to review the final draft for clarity, correctness and language. The final formulation of all results and proofs is due solely to the authors.

\section*{Acknowledgements}
The authors would like to thank Eric Ling and Jan Sbierski for identifying issues in earlier versions of (the proof of) part of \Cref{thm:main thm3} and for insightful discussions. They are also grateful to Benedikt Miethke, Miguel Prados Abad, and Roland Steinbauer for helpful discussions during the initial stages of the project.

This research was funded in part by the Austrian Science Fund (FWF) [grants DOI \href{https://www.fwf.ac.at/en/research-radar/10.55776/EFP6}{10.55776/EFP6} and \href{https://www.fwf.ac.at/en/research-radar/10.55776/STA32}{10.55776/STA32}]. KM's research was partly supported by \textbf{BASIL}, a research project funded by the European Union under the Horizon Europe programme’s Marie Skłodowska-Curie Actions Postdoctoral Fellowships [grant DOI \href{https://cordis.europa.eu/project/id/101277277}{101277277}]. For open access purposes, the authors have applied a CC BY public copyright license to any author-accepted manuscript version arising from this submission.

\printbibliography
\end{document}